\documentclass[12pt,preprint]{aastex}

\def\Msun{M$_{\odot}$}
\def\um{$\mu$m }
\def\h2o{H$_2$O}
\def\ch4{CH$_4$}
\def\arcs{\ifmmode {''}\else $''$\fi}

\begin{document}
\title{THE NIRSPEC BROWN DWARF SPECTROSCOPIC SURVEY I. LOW RESOLUTION NEAR-INFRARED SPECTRA}
\author{IAN S. MCLEAN\altaffilmark{1}, MARK R. MCGOVERN\altaffilmark{1}, ADAM J. BURGASSER\altaffilmark{1,2}, J. DAVY KIRKPATRICK\altaffilmark{3},
L. PRATO\altaffilmark{1}, SUNGSOO S. KIM\altaffilmark{4}}
\altaffiltext{1}{Department of Physics \& Astronomy, University of
California, Los Angeles, CA 90095-1562; mclean@astro.ucla.edu,
mcgovern@astro.ucla.edu, adam@astro.ucla.edu,
lprato@astro.ucla.edu} \altaffiltext{2}{Hubble Fellow}
\altaffiltext{3}{Infrared Processing and Analysis Center,
California Institute of Technology, Pasadena, CA 91125;
davy@ipac.caltech.edu} \altaffiltext{4}{Dept. of Astronomy \&
Space Science, Kyung Hee University, Yongin-shi, Kyunggi-do
449-701, Korea; sskim@ap.khu.ac.kr}

\begin{abstract}
We present the first results of a near-infrared (0.96 - 2.31
\micron) spectroscopic survey of M, L, and T dwarfs obtained with
NIRSPEC on the Keck II telescope. Our new survey has a resolving
power of R $=\lambda/\Delta\lambda$ $\sim 2000$ and is comprised
of two major data sets: 53 $J$-band (1.14 - 1.36 \micron) spectra
covering all spectral types from M6 to T8 with at least two
members in each spectral subclass (wherever possible), and 25
flux-calibrated spectra from 1.14 to 2.31 \um for most spectral
classes between M6 and T8. Sixteen of these 25 objects have
additional spectral coverage from 0.96-1.14 \um to provide overlap
with optical spectra. Spectral flux ratio indexes for prominent
molecular bands are derived and equivalent widths (EWs) for
several atomic lines are measured. We find that a combination of
four \h2o and two \ch4 band strengths can be used for spectral
classification of all these sources in the near-infrared, and that
the \h2o indexes are almost linear with spectral type from M6 to
T8. The \h2o indexes near 1.79 \um and 1.96 \um should remain
useful beyond T8. In the near-infrared a notable feature at the
boundary between the M and L types is the disappearance of
relatively weak (EW $\sim$1-2\AA) atomic lines of Al I and Ca I,
followed by Fe I around L2. At the boundary between L and T dwarfs
it is the appearance of \ch4 in all near-infrared bands ($J$, $H$
and $K$) that provides a significant spectral change, although we
find evidence of \ch4 as early as L7 in the $K$-band. The FeH
strength and the equivalent width of the K I lines are not
monotonic, but in combination with other factors provide useful
constraints on spectral type. The K I lines are sensitive to
surface gravity. The CO band strength near 2.30 \um is relatively
insensitive to spectral class. The peak calibrated flux
($F_{\lambda}$) in the 0.96 - 2.31 \um region occurs near 1.10 \um
at M6 but shifts to about 1.27 \um at T8. In addition, the
relative peak flux in the $J$, $H$ and $K$ bands is always in the
sense $J>H>K$ except around L6, where the differences are small.
One object, 2MASS2244+20 (L6.5) shows normal spectral behavior in
the optical, but has an infrared spectrum in which the peak flux
in $J$ band is {\it less} than at $H$ and $K$.
\end{abstract}


\section{INTRODUCTION}
One of the most remarkable developments in the study of stars in
recent years has been the extension of the spectral classification
sequence to low-mass objects cooler than spectral type M. The
defining characteristic of spectral class M is an optical spectrum
dominated by bands of TiO and VO, but the discovery of a cool
companion to the white dwarf GD 165 (Becklin and Zuckerman 1988)
showing none of the hallmark TiO features of M dwarfs (Kirkpatrick
et al. 1993) signaled a major change. The subsequent
identification of field counterparts to GD 165B ultimately led to
the definition of a new spectral class, the L dwarfs (Kirkpatrick
et al. 1999, hereafter K99; Martín et al. 1999), defined in the
red optical to have weakening metal-oxide bands (TiO, VO) but
strong metal hydride bands (FeH, CrH, MgH, CaH)and alkali lines
(Na I, K I, Cs I, Rb I).  To date, nearly 250 L dwarfs have been
identified\footnote{A current list is maintained by J.\ D.\
Kirkpatrick at \url{
http://spider.ipac.caltech.edu/staff/davy/ARCHIVE}}, most by
wide-field surveys: the Two Micron All Sky Survey (Skrutskie et
al.\ 1997; hereafter 2MASS), the Deep Near Infrared Survey of the
Southern Sky (Epchtein et al.\ 1997; hereafter DENIS), and the
Sloan Digital Sky Survey (York et al.\ 2000; hereafter SDSS).

As GD 165B is the prototype of the L dwarfs, Gl 229B (Nakajima et
al.\ 1995) is the prototype of a second new spectral class, the T
dwarfs. Whereas near-infrared (NIR) spectra of L dwarfs show
strong bands of H$_2$O and CO, the NIR spectrum of Gl 229B is
dominated by absorption bands from CH$_4$ (Oppenheimer et al.\
1995), features that until recently were only found in the giant
planets of the solar system and Titan. CH$_4$, H$_2$O, and H$_2$
collision-induced absorption (CIA; Saumon et al.\ 1994) give Gl
229B blue near-infrared colors, $J$-$K$ $\approx$ $-0.1$ (Leggett
et al.\ 1999). Its steeply sloped red optical spectrum also lacks
the FeH and CrH bands that characterize L dwarfs (Oppenheimer et
al.\ 1998) and instead is influenced by exceptionally broad
absorption features from the alkali metals Na and K (Burrows,
Marley \& Sharp 2000; Liebert et al.\ 2000). These differences led
K99 to propose the T spectral class for objects exhibiting $H$-
and $K$-band CH$_4$ absorption. Over 30 T dwarfs are now known.
NIR classification schemes for T dwarfs have recently been
developed by Burgasser et al.\ (2002b; hereafter B02) and Geballe
et al.\ (2002; hereafter G02). Theory suggests that L dwarfs are a
mixture of very low-mass stars and sub-stellar objects (brown
dwarfs), whereas the T dwarf class is composed entirely of brown
dwarfs, {\it i.e.}, objects with mass below that required for
stable hydrogen burning, $\sim$0.072 \Msun~ for solar metallicity
(Burrows et al.\ 2001; Allard et al. 2001).

The majority of flux ($F_{\lambda}$, in units of
Wm$^{-2}$\um$^{-1}$) emitted by L and T dwarfs is in the 1-2.5
$\micron$ NIR range, where as a result both high signal-to-noise
and high resolution studies of these intrinsically faint objects
are possible. The decreasing temperatures of late M, L, and T
dwarfs results in a rich near-infrared spectrum containing a wide
variety of features, from relatively narrow lines of neutral
atomic species to broad molecular bands, all of which have
different dependencies on temperature, gravity, and metallicity.
Furthermore, these low temperature conditions favor condensation
out of the gas state and the formation of grains ({\it e.g.,}
Lodders et al.\ 2002).

Investigations of the NIR spectral properties of M, L, and T
dwarfs have been reported previously by Jones et al.\ (1994),
Leggett et al.\ (2000, 2001), Reid et al.\ (2001), Testi et al.\
(2001), B02, and G02. Most of these observations have been carried
out at low to moderate spectral resolution (R $\sim150 - 900$) and
over a wide range of signal-to-noise. Whereas low resolutions are
appropriate for general spectral classification, higher resolution
is advantageous for detailed spectral analysis. With the advent of
new infrared detector technology, we undertook a more uniform
study of the spectra of low-mass stars and brown dwarfs, called
the Brown Dwarf Spectroscopic Survey (BDSS). The BDSS was
initiated following the successful first light commissioning of
the Keck-II Near-Infrared Spectrometer (NIRSPEC; McLean et al.\
1998, 2000a) in 1999 April. The goals of the BDSS are: (1) to
obtain a uniform set of moderate resolution (R $\sim$ 2000) NIR
spectra for a large sample of very low-mass stars and brown dwarfs
to examine general spectral properties and facilitate comparisons
with theoretical spectral energy distributions; (2) to obtain a
set of high resolution spectra (R $\sim$ 20,000) for detailed
comparison of individual spectral features with model atmospheres;
and (3) to monitor selected sources for Doppler shifts induced by
unresolved binary companions. Early products from the BDSS have
been presented in McLean et al.\ (2000b, 2001, 2003a).

This first paper in a series of BDSS results describes the
extensive observations for the initial goal and presents (a) low
resolution $J$-band spectra of 53 objects, including sixteen
spectral standards defined by K99 and B02, and (b) flux-calibrated
1.15-2.31 $\micron$ spectra of 25 objects, spanning the full range
of published spectral types from M6 to T8. Sixteen objects in the
latter set have additional coverage from 0.96-1.15 \um to provide
overlap with optical spectra. In subsequent papers relating to the
BDSS we will report on a comparison between observations and model
atmospheres, describe a similar grid of spectra developed for
gravity-sensitive indicators, and present high-resolution infrared
echelle spectroscopy of brown dwarfs (McLean et al.\ 2003b; in
prep.). In $\S$2 we describe our observations and data reduction
procedures, carried out primarily using the REDSPEC data reduction
package\footnote{For REDSPEC documentation and software, see
\url{http://www2.keck.hawaii.edu/inst/nirspec/redspec/}}. In $\S$3
we show the results in detail, analyzing the various features
present in the spectra and their evolution with spectral type.
Spectral ratios useful for classification of late-type dwarfs in
the NIR are presented in $\S$4, and compared to other
classification schemes currently defined. A discussion appears in
$\S$5. Conclusions are summarized in $\S$6.

\section{OBSERVATIONS}

\subsection{\it The NIRSPEC Instrument}

All of the observations reported here were obtained with NIRSPEC,
a cross-dispersed, cryogenic, echelle spectrometer for the
wavelength region from 0.95-5.5 \micron. Details of the design and
performance of NIRSPEC are given elsewhere (McLean et al. 1998,
2000a). A low-resolution setup of R $\sim$ 2000 for the same slit
width of $\sim$ 0.4$\arcsec$ (only 2 pixels in this mode) is
available without the echelle grating. This is the mode we have
employed for the data presented here. Dispersion along the
detector is approximately 2.5 \AA/pixel at 1 \micron. NIRSPEC has
3 detectors: a sensitive CCD camera for acquisition and offset
guiding, a slit-viewing camera with a 256 $\times$ 256 HgCdTe
array sensitive from 1-2.5 \um, and the primary 1024 $\times$ 1024
InSb array for the spectrograph. The InSb array has a high quantum
efficiency $(\sim80$\%) across the 1-5 \um region with a dark
current of 0.2 $e^-$ s$^{-1}$ pixel$^{-1}$ at 30 K and a noise of
$\sim$ 15 $e^-$ rms with multiple non-destructive readouts.
Typical sensitivity values under good seeing conditions in
low-resolution mode yield signal-to-noise ratios of 25 per
resolution element at $H \sim 15.5$ in 20 minutes. NIRSPEC remains
stationary on kinematic mounts on the Nasmyth platform of the Keck
II telescope and operates at the f/15 focus with the tertiary
mirror in place. The optical properties of the high-transmission
Three-Mirror-Anastigmat (TMA) camera in NIRSPEC necessitates
certain rectification requirements during data reduction.

For low-resolution spectral observations, seven custom-designed
order-sorting filters, referred to as NIRSPEC-1 through NIRSPEC-7,
give overlapping wavelength coverage from 0.95 - 2.6 \micron.
Table 1 lists the six instrumental configurations we have used.
The filter, the grating angle in the low resolution mode, the
wavelength coverage, and the corresponding photometric band are
listed. The NIRSPEC-1 filter covers the optical and near-infrared
$z$-bands (e.g., Gunn-$z$, UFTI $Z$), as well as the Y-band
(Hillenbrand et al.\ 2002). NIRSPEC-2 bridges the $z$ and
$J$-bands; the NIRSPEC-3 filter covers the $J$-band. NIRSPEC-4
spans the short wavelength half of the $H$-band. NIRSPEC-6 is a
broad $H$+$K$ filter centered at 1.925 \um with a bandwidth of
0.75 \micron. Since the entire NIRSPEC-6 bandwidth cannot be
simultaneously observed on the detector, two configurations, N6a
and N6b, are used to obtain the long wavelength half of the
$H$-band and the $K$-band, respectively. NIRSPEC-5 (essentially
$H$-band) and NIRSPEC-7, which cover 1.431 - 1.808 \um and  1.839
- 2.630 \micron, respectively, were used only for one NIRSPEC-5
observation of the T8 dwarf 2MASSW J0415-09, and NIRSPEC-7
observations of the L5 dwarf 2MASS W J1507-16 and the L8 dwarf
2MASSW J1523+30. While NIRSPEC-7 extends wavelength coverage well
beyond the 2.295 \um CO band, its use was deemed too costly in
observing time (because of high backgrounds and poor
signal-to-noise) to include in the survey.

\subsection{\it Observing Strategy}

Table 2 lists the sample of low mass stars and brown dwarfs
included in our survey. Coordinates (J2000), 2MASS photometry,
published spectral types and the date of the observation are
included for reference. For convenience, the full 2MASS names are
truncated but no confusion arises because there are no duplicates
and the full J2000 coordinates (which define the name) are given.
The apparent magnitudes of the sources are in the range 7 $<$ $J$
$<$ 16.8. Targets were selected from well known M dwarfs and from
L and T dwarfs identified in 2MASS (Kirkpatrick et al.\ 1997, K99,
2000, 2001; Burgasser et al.\ 1999, 2000a, 2000b, B02; Reid et
al.\ 2000; Wilson et al.\ 2001; Dahn et al.\ 2002; C.G. Tinney \&
J.D. Kirkpatrick 2003, private communication), augmented with
discoveries from DENIS (Delfosse et al. 1997), SDSS (Strauss et
al. 1999; Leggett et al. 2000; G02), and other investigations
(Becklin \& Zuckerman 1988; Ruiz, Leggett, \& Allard 1997). Our
primary goal was to obtain a complete sample of $J$-band spectra
from M6 - T8 with at least two representatives in every spectral
class if possible, followed by spectra over the full 0.96-2.315
\um range for all even-numbered spectral subclasses.

For most of these objects the same observing strategy was
employed. A 300s exposure was obtained with the source at each of
two positions (called nod positions) separated by $\sim$20\arcsec~
along the 0$\farcs$38 wide slit. Most objects were observed in an
ABBA pattern giving a total integration time of 20 minutes per
grating/filter configuration. Six settings (2 hr) were required to
obtain a complete spectrum from 0.960-2.315 \micron~ that was
continuous through the NIR atmospheric absorption bands. Shorter
exposures were used for the brightest sources to avoid detector
non-linearity or saturation. Total integration times were doubled
(40 minutes per grating setting) to maintain a signal-to-noise
ratio of at least 20:1 per resolution element over most of the
filter band in the faintest sources ($J \sim $16). Sources with $J
\leq$ 14 have a signal-to-noise ratio over 100:1 in 20 minutes.
Prior to early 2000, we used longer on-chip exposure times of 600s
for some objects, but discontinued this practice when some OH
night-sky lines were found to be saturating (especially in the $H$
band) and not subtracting out properly in the data reduction
process (see McLean et al.\ 2001). Seeing was generally good on
most nights ($\sim$0.5\arcs), with only a few nights worse than
0.7\arcs-0.9\arcs. The same slit width was maintained throughout.

To account for absorption by the Earth's atmosphere, calibration
stars of spectral type B9V-A3V were observed before or after the
target observations and as close to the same airmass and time as
possible, typically within 0.1 airmasses but in a few cases as
high as 0.3. Early A type stars are essentially featureless in
$JHK$ except for hydrogen absorption lines, which can be
successfully interpolated out during the data reduction process.
The $J$ and $K$ bands each contain one line, Pa$\beta$ at 1.282
\micron~ and Br$\gamma$ at 2.166 \micron. The $H$ band contains
Brackett lines at 1.737, 1.681, 1.641, 1.611, 1.588, 1.571, 1.556,
1.544 and 1.534 \micron. Both neon and argon arc lamp spectra were
obtained either immediately before or after the target observation
for wavelength calibration, together with the spectrum of a
flat-field lamp and a dark frame.

\subsection{\it Data Reduction}

Reduction of the spectra was accomplished using REDSPEC, software
designed and developed primarily for NIRSPEC data reduction. At
each step in the procedure graphical displays allow inspection of
the process by the user. As a result of the high-throughput
optical design ($\S$ 2.1), NIRSPEC spectra are non-linear in the
spatial and spectral dimensions. REDSPEC rectifies the data,
checks for saturation and performs standard reduction techniques.
The code first maps the spatial distortion by fitting a polynomial
to the location of the data across the two-dimensional array. A
spectral map is then created by fitting a polynomial to observed
neon and argon arc lamp lines with known
wavelengths\footnote{Wavelengths of arc lamp lines may be found in
the NIST database, \url{http://physics.nist.gov/}}, producing a
dispersion solution for every row of (spatially rectified) data.

After rectification, the pairs of nodded target frames are
subtracted to remove background and then divided by the rectified
flat field. Image frames are cleaned by replacing bad pixels by
interpolating over their neighbors. Comparison of the target
spectrum with the OH night-sky spectrum is used to reveal any
artifacts caused by incomplete subtraction and saturated OH lines.
Both target and calibrator spectra are extracted by summing
$\sim$10 adjacent rows of data. After interpolation across
intrinsic stellar features, the calibration star spectra are
divided into the corresponding target spectra to remove telluric
absorption features and instrument spectral response. The
resulting spectra are subsequently multiplied by a blackbody
curve, equivalent in temperature to that of the calibration star
(Tokunaga 2000), to restore spectral slope and permit the
calculation of accurate fluxes. At this stage the spectra
extracted from the two nod positions are averaged, and the action
of changing the sign of the negative spectrum and adding to the
positive spectrum eliminates any remaining sky offset.

The spectra were flux calibrated (in $F_{\lambda}$ units)using
2MASS photometry, following the prescription of McLean et al.\
(2001). Briefly, we compared the band-averaged flux densities for
each target with those of Vega and derived a calibration factor by
matching to the known magnitude in that band. Code was developed
to bootstrap the overlapping spectral pieces into a continuous
spectrum while minimizing photometric errors. Comparison of the
bootstrapping method with direct flux calibration in individual
wavebands yields photometric errors consistent with 2MASS
magnitude uncertainties. Our $F^{band}_\lambda ({\rm Vega})$
values are based on data from Bergeron, Wesamael, \& Beauchamp
(1995), kindly provided by D. Saumon, and the 2MASS filter set. We
also compared these values to fluxes from Cohen et al. (1992),
which are based on the UKIRT filter set and the atmospheric
absorption at Kitt Peak (Cox 2000). A 5\% difference was found in
the $J$-band, but only 1-2\% difference in the $H$ and $\rm K_s$
bands.

\section{RESULTS: MORPHOLOGY OF THE SPECTRA}

Figures 1 and 2 present 39 $J$-band spectra ordered with respect
to their optical (L dwarfs) or infrared (T dwarfs) classifications
given by K99 and B02 respectively. We will discuss in $\S$4
whether or not this order is consistent with a pure infrared
classification scheme. Objects range from the M6 dwarf Wolf 359 to
the T8 dwarf 2MASS 0415-09, the coolest brown dwarf yet discovered
(B02; F. Vrba et al. 2003, private communication). There are no L9
objects in the K99 classification scheme and a T4 object is
missing from the sequence as no such object has yet been
identified. There are two objects of each published spectral type
wherever possible. These 39 spectra are a subset of the 53 spectra
that are included in the initial phase of the NIRSPEC survey. The
remaining 14 objects are either non-integer spectral class
designations, redundant with the included spectral classes, or
peculiar and will be presented in $\S$3.5. Figure 3 shows a
sequence of 12 complete flux-calibrated spectra ($F_{\lambda}$ vs.
$\lambda$) for every even-numbered spectral type from M6 to T8,
with a T5 spectrum substituted for the missing T4. Spectral
coverage is continuous from 0.96 - 2.315 \micron. Noisy regions at
1.11-1.14 \um, 1.35-1.45 \um and 1.82-1.95 \um correspond to
terrestrial atmospheric absorption bands that separate and limit
the $zJHK$ bands. These flux-calibrated spectra have been
normalized to the $J$-band at 1.27 \um for plotting. For
reference, Table 3 gives the normalization fluxes (F$_{\lambda}$)
at 1.27 $\mu$m.

Figures 1-3 show a rich variety of atomic and molecular spectral
features that clearly evolve with spectral type. Essentially all
of the fine spectral structure is real and repeatable, except in
regions of very poor atmospheric transmission. Before
quantitatively analyzing the entire data set, it is useful to
review the trends and features qualitatively. To illustrate the
intrinsic spectral resolution of the NIRSPEC BDSS, Figures 4-7
provide close-up plots of five representative objects in the $z,
J, H$ and $K$ spectral bands: Wolf 359 (M6), 2MASS 0015+35 (L2),
Gl337C (L8), SDSS 1254-01 (T2), and Gl570D (T8). Below, we
describe the spectral morphology in each band.

\subsection{\it $z$-band}

Figure 4 contains the $z$-band spectra of our five representative
objects. The most prominent feature of this region is the
well-known Wing-Ford band (Wing \& Ford 1969) of FeH. The band
head at 0.9896 \micron~ comes from the 0-0 ($v' - v''$) transition
of the F$^{4}\Delta$--X$^{4}\Delta$ system. This feature
strengthens from M6 through the early L dwarfs and then fades
through the late L dwarfs and early T dwarfs (see also Figure 2),
but then strengthens again from T4-T6 before vanishing at T8. The
0.9896 \um band head is accompanied by many other transitions of
FeH as demonstrated by Cushing et al.\ (2003). Those authors
identified 33 absorption features redward of this band head (from
0.9975-1.0849 \um) as FeH bands. Each of these features is clearly
seen in the spectra of our early L dwarfs, including the weaker of
two Q-branches at 0.9979 \micron. The feature at 0.9969 \um has
generally been attributed to the 0-1 A$^{6}$ - X$^{6}$ band head
of CrH (K99), although Cushing et al.\ (2003) have cast doubt on
this identification, attributing absorption in this region
entirely to FeH. Recent opacity calculations for CrH (Burrows et
al.\ 2002) and FeH (Dulick et al.\ 2003) confirm the presence of
the CrH band, but its observability depends strongly on the
relative abundance of FeH/CrH in these cool atmospheres, which is
currently unknown. The broad shallow feature at 1.0666 \um is
unidentified, but behaves in the same way with spectral type as
the FeH and CrH features. In the L sequence the continuum begins
to slope downwards towards the blue end of this band and the \h2o
feature near 1.11 \um strengthens. This trend continues into the T
dwarfs, until, by T8, the spectrum is relatively smooth and
strongly peaked near 1.08 \micron.

\subsection{\it $J$-band}

Figure 5 is a plot of the extended $J$-band region where we have
combined data from the N2 and N3 filters. In the spectrum of the
M6 dwarf there is a strong neutral sodium (Na I) doublet (1.138
and 1.141 \micron) and a weaker Na I line at 1.268 \um. There are
also two pairs of neutral potassium (K I) doublets (1.168,1.177
\um and 1.243,1.254 \micron), a doublet of Al I (1.311, 1.314
\micron), two prominent Fe I lines (1.189 and 1.197 \micron), and
three weaker Fe I lines at 1.1596, 1.1610 and 1.1641 \micron. Very
weak features from Ti I and Mn I at 1.283 and 1.290 \um
respectively, and two sets of Ca I triplets are not convincingly
detected after M6. The prominent metal lines weaken and disappear
near the M/L boundary (McLean et al. 2000b). The sodium and
potassium doublets persist through most of the L sequence, after
which the Na doublet weakens markedly. The K I doublets,
especially the long wavelength pair of lines, can be traced until
T7, but are exceedingly shallow or absent at T8. The K I lines
also become consistently broader with later spectral type.
Absorption by \h2o occurs at both ends of this spectral region.

The narrow absorption feature at 1.135 \um is from \h2o, and
weaker features, most notable in the L dwarfs, spread redward to
1.195 \micron. A strong \h2o band occurs longward of 1.33 \micron.
In late-type M stars this \h2o band can depress the flux at 1.34
\um by 10-20\% from the apparent continuum level. Lying between
the two K I doublets are FeH bands in the range of 1.19--1.24
\micron. Cushing et al.\ (2003) identify the prominent features at
1.1939 and 1.2389 \um as the band heads of the 0-1 and 1-2 bands
of the F$^{4}\Delta$-- X$^{4}\Delta$ system. These FeH features
strengthen in the L dwarfs until about L5, after which they decay
and are absent in the T dwarfs. The \h2o absorption band longward
of 1.33 \micron~ increases in strength through the L dwarf
sequence, diminishing the continuum by 30-40\%.

In the late L dwarfs (e.g., Gl337C), the flux between 1.28 and
1.32 \um begins to slope slightly downward as a result of
increasing \ch4 absorption. This absorption becomes a clear
feature in the T dwarfs as the band strengthens. Blueward of
$\sim$1.19 \um, \h2o and \ch4 absorption bands cause a decrease in
flux which continues throughout the T dwarf sequence.

\subsection{\it $H$-band}

Figure 6 shows the 1.36-1.93 \um $H$-band region. Again, two
NIRSPEC settings are combined here to extend the region beyond the
standard $H$-band, which explains the dense appearance of the
spectra. Telluric H$_{2}$O absorption, even from a high-altitude
site like Mauna Kea, contaminates the wavelength regions from
1.34-1.41 \micron~ and 1.80-1.96 \micron, causing some spectra to
exhibit spurious features. Most of the $H$-band spectra are
dominated by very broad H$_2$O absorption intrinsic to the sources
around 1.4 and 1.8 \micron.  These \h2o bands strengthen
significantly from the M dwarfs to the T dwarfs; by spectral type
T8, water vapor absorption has removed most of the flux in the
range 1.45-1.50 \micron.

One feature which is clearly present in the late M and early L
dwarfs is the blended K I doublet line at 1.517 \micron. This
feature can be seen in 2MASS 0015+35 (L2) but is no longer present
in the L8 object, Gl 337C. Cushing et al.\ (2003) have identified
34 features in this spectral region belonging to the 0-0 band of
the E$^{4}\Pi$-A$^{4}\Pi$ system in FeH. Band heads are present at
1.5826, 1.5912 and 1.6246 \micron, labelled in the figure. FeH
weakens through the L dwarf sequence. The onset of the 2$\nu_{3}$
band of CH$_{4}$ at 1.67 \micron~ can be seen in the L8 Gl337C,
but is much clearer in the T2 dwarf. This feature strengthens
dramatically through the T dwarf sequence in addition to the
$2\nu_{2}+\nu_{3}$ band at 1.63 \micron, clearly distinguishing
these objects.

\subsection{\it $K$-band}

Figure 7 shows the enlarged $K$-band spectral region for the same
five objects. Atomic features of Ca I (1.98 \micron~ triplet) and
Na I (2.206,2.209 \micron) are present in the M dwarfs but
disappear in the early L dwarfs. The CO (v=2-0) band head at 2.295
\um is also present in the M dwarfs, increases slightly in
strength in the early L dwarfs, but weakens again in later types.
\h2o absorption around 1.8 -2.0 \um is weak in the M dwarfs but
appears in the early L dwarfs and gets stronger through the L and
T dwarf sequence.  An additional source of continuum opacity in
this region is collision induced absorption (CIA) by H$_2$
molecules (Saumon et al.\ 1994) which has no distinct bandhead but
suppresses flux throughout the $K$-band. Absorption from the
$\nu_{2}+\nu_{3}$ band of \ch4 at 2.2 \um appears in the latest L
dwarfs (McLean et al.\ 2001, Nakajima et al.\ 2001, G02), and like
its $J$- and $H$-band counterparts, strengthens in the T dwarfs.
CO is still detectable in the T2 dwarf, SDSS 1254-01, but does not
appear in the late T dwarfs.

\subsection{\it Additional Spectra}
Figures 8 and 9 show the remaining $J$-band spectra in the survey.
Several of these objects have redundant spectral types or are
objects which have been assigned half-class designations. For one
spectral type, L1, we have six objects. These are compared in
Figure 9 and appear to show good agreement.

Figure 10 shows an additional group of 12 objects for which
flux-calibrated spectra have been obtained from 1.15 - 2.315 \um.
With the exception of four objects, the region from 0.96-1.15 \um
linking the infrared spectrum to the far-red optical is missing.
In addition, the T8 2MASS 0415-09 does not have continuous
coverage through the atmospheric absorption bands. Some of these
sources are faint and the spectra are noisier in the terrestrial
atmospheric bands.

\section{ANALYSIS}

Molecular absorption bands constitute the most prominent features
in the near-infrared spectra of M, L and T dwarfs. Indeed, most of
the NIR classification studies to date are based largely on the
strengths of the \h2o and \ch4 features. In this section we
examine and quantify the evolution of these strong, broad features
as well as other molecular and atomic features.

\subsection{\it Spectral Flux Ratios for Molecular Bands}

Following earlier work on infrared spectral indexes by Jones et
al.\ 1994; Tokunaga \& Kobayashi 1999; Reid et al.\ 2001; McLean
et al.\ 2000b; Testi et al.\ 2001; B02; G02, we define a set of
flux ratios for NIRSPEC spectra in the primary \h2o, \ch4, FeH,
and CO bands.

Table 4 lists the flux ratios used and provides their definitions.
Nine ratios have been created to measure the most prominent
molecular features in the $zJHK$ bands; four \h2o bands (1.343
\micron, 1.456 \micron, 1.788 \micron, 1.964 \micron) labelled A,
B, C and D respectively, two \ch4 bands (1.730 \micron, 2.2
\micron) labelled \ch4A and \ch4B respectively, two FeH bands
(0.988 \micron, 1.200 \micron) designated $z$-FeH and $J$-FeH, and
the primary CO band (2.295 \micron). Each ratio is formed by
determining the median in a 0.004 \micron~ (40 \AA) wide interval
at the wavelength of the feature and dividing this by the median
continuum level over the same interval; the greater the
absorption, the smaller the ratio. Because the equivalent
resolving power per pixel corresponding to our 40 \AA bandwidth
ranges from R = 250-570 (depending on wavelength), our ratios
could also be relevant for lower resolution spectroscopy, although
with fewer pixels per band. Many possible center wavelengths,
continuum wavelengths and bandwidths were analyzed for each ratio
before making a choice that maximized sensitivity to both the
given spectral feature and the linear response.

These ratios, which serve to quantify the NIRSPEC BDSS data set,
can be compared to prior spectral classifications, especially
optical types, to investigate if the same underlying physics is
controlling both the optical and NIR spectral signatures. Table 5
lists the values of these nine flux ratios for the objects in the
survey. Figures 11 - 13 contain plots of these values as a
function of published optical spectral subclass for L dwarfs or
previous NIR class for T dwarfs. Filled symbols in these plots
identify published spectral type standards from K99, B02 and G02.
Some ratios are well-defined only for certain spectral classes.
For example, the four \h2o ratios are defined for all subclasses
from late M to late T since \h2o absorption plays a significant
role in the spectral energy distribution of all these objects. The
\ch4 ratios, however, are defined only for T dwarfs, as \ch4
absorption is absent in the M and early L dwarfs, although the
onset of \ch4 can be deduced from the values of these ratios for
the late L dwarfs. The FeH ratios are only well-defined for M and
L dwarfs, although the 0.988 \micron~ band can be seen in some of
the T dwarfs. Finally, the CO ratio is useful only for objects in
the range M6 to $\sim$T3 where the CO band head is clearly
identifiable.

Examination of Figures 11-13 reveals that, in several cases, the
behavior of the selected flux ratios is monotonic with prior
spectral type assignments and approximately linear. Where
possible, the best linear fit to the indexes is shown together
with a $\pm$1 $\sigma$ error bar derived from the residuals to the
line fit. Table 6 lists the straight line fit parameters (slope
and intercept), as well as the correlation coefficient ($R^2$) for
the least-squares solution and the standard deviation of the
scatter in spectral type. We adopt the convention that M0 = 0, L0
= 10 and T0 = 20.

Comparing \h2o ratios, it appears that the \h2oA index, although
monotonic, is not well fit by the same line in the T dwarfs
because of contamination by \ch4 at 1.313 \um. A decrease in the
continuum level at 1.313 \um drives the \h2oA ratio higher. The
effect is minor but nevertheless evident in Figure 12. Before L6,
the data are evenly spread about the line while beyond L6 the data
could be better fit with a slightly steeper line or a weak second
order polynomial. Of the four \h2o indexes, \h2oB seems to show
the best linear correlation with previously assigned spectral type
across the entire sequence. The \h2oC ratio exhibits behavior
similar to the \h2oA index because the onset of \ch4 absorption in
the T dwarfs influences the continuum level. Again, a weak
spectral break to a slightly steeper linear fit occurs around L6.
In the $K$-band, the \h2oD ratio appears to have a good linear
correlation with spectral type.

Our \ch4 indexes have an excellent linear correlation with the
near-infrared derived T dwarf spectral types assigned by B02 and
G02. Excluding the last two data points in the \ch4A and \ch4B
plots in Figure 12 (the T8 dwarfs Gl570D and 2MASS 0415-09)
improves the fit slightly, most likely as a result of saturation
of the \ch4 bands.

The behavior of the FeH and CO indexes is more complex. Both of
the FeH indexes decrease in value as the band strength increases
until L3.5-L4, where they reach minima and account for a 25\% drop
in the continuum in the $J$-band and a 60\% drop in the $z$-band.
An abrupt change in the FeH flux ratios can be seen at the
transition from L5 to L6, possibly related to peaks in the \h2oA
and \h2oC indexes at the same spectral types. At later spectral
types the index increases steadily to spectral type T5.5 because,
in both $z$ and $J$ it is no longer measuring FeH but the slope of
the continuum. Very weak $J$-band FeH can be seen in the T5 2MASSW
0559-14, although its relatively bright $J$ magnitude of 13.83
makes the detection of subtle features easier than in other
fainter T dwarfs. However, the T2 object SDSS 1254-01 has no
discernable FeH band at 1.14 \micron, whereas the 0.988 \micron~
FeH band in $z$ is detected as late as T6. Beyond T6 the ratio
rises steeply as a result of the influence of \h2o absorption on
the continuum and the index no longer has any physical meaning
with regard to FeH absorption.

The CO flux ratio (Fig.13) shows little variation from M6 to T2.
There may be a slight minimum in the mid-L dwarfs, but the data
deviate from a constant by no more than 15\%, similar to the
result found by Reid et al.\ (2001). Later than T2 the ratio has
no meaning for CO and would show a (noisy) rising trend that grows
in value as \ch4 absorption depresses the continuum relative to
the wavelength of the CO band.

\subsection{\it Alkali Line Equivalent Widths}

The dominant atomic line features in the near-infrared are the
four neutral K I lines in the $J$-band. Table 7 gives the
calculated equivalent widths and estimated errors in \AA, while
Figures 14 and 15 show the variation of equivalent width for each
K I doublet as a function of published spectral type. For the
short-wavelength pair, the best estimate of the local continuum
was calculated by finding the median value in a 0.002 \um window
on both sides of each line; the continuum point between the lines
was usually common to both lines. The selected wavelengths were
1.162 \micron, 1.1735 \micron~ and 1.185 \micron. For the
long-wavelength pair the continuum points were centered at 1.233
\micron, 1.248 \micron~ and 1.260 \micron. From these points, a
linear interpolation was made to determine the approximate
continuum across the line. Equivalent widths were measured by
summing the residual intensities (interpolated continuum minus
line) between specific wavelength intervals and multiplying by the
resolution element in angstroms. The limits used were: 1.167-1.171
\micron, 1.175-1.180 \micron, 1.2415-1.246 \micron~ and
1.2498-1.2558 \micron. The choice of 1.2415 \um for the 1.2435 \um
K I line is to avoid contamination by the FeH band at 1.24
\micron. We visually confirmed that these wavelength regions were
appropriate for each spectrum. To determine an uncertainty for the
equivalent width, the location of the continuum reference points
was allowed to vary by about 10\% of the sample bandwidth and the
value of equivalent width was recalculated. The tabulated
equivalent width is the median of 25 such trials and the
uncertainty is the standard deviation in this set.

Although there is considerable scatter, general trends can be seen
in Figures 14 and 15. The K I lines strengthen across the M to L
dwarf boundary, form a broad peak around L4 or L5, decline towards
the late L types, and rise again at the L/T boundary before
finally disappearing in the latest T dwarfs. The longer wavelength
pair can be followed to T6. We also found that the K I lines are
narrower and weaker in very young low-mass objects with lower
surface gravity. Those results will be reported elsewhere
(McGovern et al. 2003, in prep.).

At 1.1385 and 1.1407 \um there is a relatively strong Na I feature
which, in an M6 dwarf, is even stronger than the K I doublets. As
shown by the variation in equivalent width of the doublet in Fig.\
16, this Na I doublet strengthens from M6 to a broad peak around
L3/L4, weakens rapidly through L8 and is absent by T0. Another
weaker Na I doublet occurs at 2.2 \um and is quite prominent from
M6-M9 but does not persist past L2. The variation in equivalent
width of this doublet is shown in Fig.\ 17 and tabulated in Table
8.

\subsection{\it Weak Atomic Features of Fe, Al and Ca}
Despite the extensive blanketing of the $JHK$ region with
molecular transitions, several interesting atomic lines of more
refractive elements can be studied. Figures 18 and 19 show the $J$
and $K$ regions for an M8, M9, L0 and L1 dwarf. The Al I doublet
in the $J$-band disappears at the M9/L0 transition. Although
heavily contaminated by \h2o absorption, the $K$-band Ca I triplet
seems to persist to $\sim$L2. This transitional change at the M/L
boundary is also seen in the equivalent widths of the Al I doublet
(Figure 20). Because of the \h2o contamination equivalent widths
are difficult to measure for the Ca I features and none are
reported at this time. There are also several Fe I lines in the
$J$-band, but the least contaminated feature is the line at 1.189
\um which has an equivalent width of about 1 \AA~ at M9. As shown
in Fig.\ 21, this line persists until at least L3 before
disappearing. Table 8 includes the equivalent width of the Al and
Fe lines along with the Na lines mentioned above.

\subsection{\it Spectral Energy Distributions}
The behavior of the flux-calibrated ($F_{\lambda}$) spectra given
in Figures 4 and 10 can be quantified. Figure 22 shows the ratio
of the peak calibrated flux in the $H$- and $K$-bands relative to
that in $J$ derived from the spectra. For convenience of plotting,
the calibrated flux spectra were normalized using the mean flux
value in $J$-band in the wavelength region 1.24 - 1.29 \micron.
The peak flux of the $H$- and $K$-bands were measured from the
mean fluxes in the wavelength regions 1.55 - 1.60 \um and 2.08 -
2.13 \um respectively, and compared to the peak value in the
$J$-band. At all spectral types, the peak flux at $J$ exceeds that
at $H$ or $K$, but there is a notable trend through the mid-L
dwarfs in which the relative difference weakens until the fluxes
are almost equal, and then the $J$ band recovers in the T dwarfs.
This spectral result is of course mirrored in the $JHK$ colors and
is generally attributed to the formation, and settling out, of
dust (Tsuji 2001; Ackerman \& Marley 2001).

\section{DISCUSSION}
\subsection{\it Fits to \h2o Ratios}
In presenting the NIRSPEC BDSS results in this paper we have used
the published L spectral types from Kirkpatrick et al.\ (1999,
2000), and therefore the classification scheme described in K99.
Since much of the energy emission occurs in the near-infrared, it
is important to determine if the optically-defined classification
scheme for L dwarfs yields an ordering that is also consistent
with observed spectroscopic morphology in the near-infrared. At
present, there is no agreed classification scheme. Visual
inspection of the 53 $J$-band spectra, using the apparent growing
strength of the \h2oA absorption band as the primary guide for M -
L dwarfs, reveals a remarkable consistency with far red optical
classification. A similar conclusion can be reached on examining
the set of 25 spectral energy distributions extending wavelength
coverage out to 2.315 \um. Given the large wavelength range
involved, this is an encouraging result. This overall impression
is strengthened by examination of flux ratio plots (Figs.11-13) as
a function of optical spectral type. Note that the objects plotted
as solid black points are the previously defined standards. From
the best fitting lines to the four \h2o indexes it is possible to
write expressions that allow spectral types to be predicted from
the NIRSPEC flux ratios. These expressions are:
\begin{center}
\begin{equation}\label{1}
    Sp= -26.18(H_{2}OA)+28.09~~~\sigma(Sp)=\pm 1.1
\end{equation}
\begin{equation}\label{2}
    Sp= -22.94(H_{2}OB)+29.54~~~\sigma(Sp)=\pm 0.6
\end{equation}
\begin{equation}\label{3}
    Sp= -39.37(H_{2}OC)+38.94~~~\sigma(Sp)=\pm 1.8
\end{equation}
\begin{equation}\label{4}
    Sp= -25.06(H_{2}OD)+34.51~~~\sigma(Sp)=\pm 0.8
\end{equation}
\end{center}

We are not advocating a spectral classification scheme based
entirely on a single molecular species. These relationships
demonstrate however, that a strong and useful correlation exists.
When all four \h2o ratios are available, a weighted mean using the
1$\sigma$ errors as weights, gives the best overall fit to the
existing spectral classes. Of course, the derived spectral class
must also be consistent with the FeH and alkali line strengths,
and this is generally the case. In applying these relationships to
NIRSPEC observations, the \h2o ratios serve to identify the likely
spectral type. Classification then follows by a detailed
comparison of the spectrum to that of the standard for that type.
Nevertheless, there are discrepancies and limitations to the
correspondence between the optical and NIR spectral types. For
example, from the six L1 dwarfs observed in $J$-band (Fig. 9) the
average \h2oA flux ratio is 0.633$\pm$0.040 (6.25\%), which
translates to an error of $\pm$1.1 in spectral type, consistent
with the above linear relations. It is curious that 2MASS 1300+19
has the highest equivalent widths in K I of any L1, or indeed
almost any other L type. On the basis of the \h2oA index and the K
I equivalent widths, the infrared type might be as late as L3.
Also, as shown in Figs. 14 and 15, the L2 dwarf 2MASS 1726+15 and
the L6.5 dwarf 2MASS 2244+20 have weaker K I lines than expected
from the trend. In addition, Kelu-1 (L2) also appears weaker than
might be expected. These effects are possibly related to gravity
and/or metallicity rather than to effective temperature, with the
caveat that linear fits to \h2o indexes need not be linear with
effective temperature. We will shortly report on a study that
demonstrates that younger objects exhibit weaker alkali lines
(McGovern et al. 2003; in prep.)

\subsection{\it Comparison with other classification schemes for L dwarfs}
Following the K99 classification scheme, Figure 2 does not contain
any objects classified as type L9. G02 developed a near-infrared
scheme optimized for T dwarfs, with L dwarfs being classified the
same way by noting the continuity in spectral indexes. The main
difference was that some optical L8 objects were re-classified as
L9. For example, G02 assigned L9 $\pm$ 1 to 2MASS 0310+16 and L9.5
$\pm$ 1 to 2MASS 0328+23 (see Fig. 8). At least in the $J$-band,
the morphological distinction is marginal between L8 and T0,
although it can be argued that both of these L8 objects do appear
as slightly later spectral types. The $J$-band spectra of the five
L8 dwarfs in our survey are collected in Fig.\ 23. The average
value of the \h2oA index for the five L8s in our survey is
0.417$\pm$0.033 (7.9\%), which is closer to L7. 2MASS 1632+19,
classified as L8 by K99, is assigned L7.5 by G02, whereas a
weighted average of our four \h2o ratios yields L7. If the \ch4
indexes are included however, then 2MASS 1632+19 appears
intermediate between an L8 and a T0, and the final estimate would
be L8$\pm1$. Clearly, the classification of objects near the L/T
boundary requires further study.

DENIS-P J1228-15AB is classified as L5 by K99 and L6$\pm$2 by G02
indicating that its infrared signature is less conclusive. From
Fig. 1 it is clear that its FeH strength is too great for L6 and
its \h2oA band is deeper than in L4. There is also a peculiar
depression in the $J$-band spectrum near 1.29 \um and a slightly
stronger slope to shorter wavelengths. Our results suggest L5 at
$J$-band, but with peculiarities. Another curious object is
DENIS-P J0205-11AB which is classified L5.5$\pm$2 by G02 and L7 by
K99. Examination of Figs. 2, 10 and 11 strongly suggest a late
spectral type, including even the suggestion of \ch4 absorption.
Our \h2o ratios yield L8, but the \ch4 indexes suggest between L8
and T0.

\subsection{\it Comparison with other classification schemes for T dwarfs}
Published T dwarf classifications from B02 and G02 already include
near-infrared indexes, and therefore correspondence with the
NIRSPEC data should be good. The only difference is SDSS 1624+00
which is given as T6 by both B02 and G02, yet appears more like T7
at $J$-band in Figure 2, but has $H$ and $K$ \ch4 ratios
completely consistent with T6. These differences are small and
imply substantial consistency between NIR types and our linear
indexes. The following relations demonstrate the expected good
correspondence of our \ch4 indexes to published spectral type.
Again, we are not advocating that T dwarf classifications be based
on a single molecular species. A weighted combination of the \ch4
and \h2o indexes is very effective for assigning spectral type.
\begin{center}
\begin{equation}\label{5}
    Sp= -11.82(CH_{4}A)+29.42 ~~~\sigma(Sp)=\pm 0.2
\end{equation}
\begin{equation}\label{6}
    Sp= -9.94(CH_{4}B)+28.36 ~~~\sigma(Sp)=\pm 0.4
\end{equation}
\end{center}

\subsection{\it Further discussion of the BDSS}
One remarkable feature apparent in our collection of NIR spectra
of M, L and T dwarfs is the continuous increase in \h2o
absorption. Until the onset of \ch4 absorption at the L/T
boundary, changes in spectral morphology are not as apparent as in
the optical where TiO declines and gives way to CrH. Consequently,
the M/L boundary is poorly delineated in the NIR. With the
resolution of the BDSS however, a clear transition is revealed by
the behavior of the refractory elements Al and Fe. The equivalent
width of the 1.189 \um Fe I line begins to decline after M9, and
the Al I doublet disappears from the $J$-band between M9 and L0.
Interestingly, the Ca I lines in the $K$-band persist until at
least L2, and the neutral Fe line at 1.189 \um finally disappears
after L3.

The Al I doublet involves transitions between energy levels at
3.14 and 4.09 eV, whereas the Fe I line comes from states at 2.176
and 3.211 eV. The Na I doublet at 1.14 \um involves energy levels
at 2.10 and 3.19 eV; this line weakens markedly after L4. The
$J$-band K I lines are all associated with even lower energy
levels (1.61 to either 2.67 or 2.61 eV). On the other hand, the
$K$-band Ca I lines are associated with transitions that require
the population of higher level states (4.62-5.25 eV, and 3.91-4.53
eV). Because the Na I doublet at 2.206 \um involves states at
3.19-3.75 eV, one might expect the Ca I lines to vanish at an
earlier spectral type, but they do not.

This behavior may offer constraints on condensation chemistry
because, as shown by Lodders (2002), the equilibrium chemistry of
Ti, V, Ca and Al is inextricably linked together. Based on the
analysis of Lodders (2002), the highest temperature condensate
containing either Ca or Ti is hibonite (CaAl$_{12}$O$_{19}$). The
condensation temperatures depend on total pressure, but
calculations for 1 bar seem to be applicable (K. Lodders 2003,
private communication). Hibonite starts condensing at 1997 K but
is replaced by grossite (CaAl$_4$O$_7$) at 1977 K. Condensates
bearing titanium follow, and by 1818 K, where perovskite
(CaTiO$_3$) is stable, $\sim$97\% of all Ca is expected to be tied
up in solids. Because \h2o contamination in this region makes it
hard to quantify when the line is really gone, 1818 K is a
conservative lower limit. If there is still as much as 10\% of all
Ca in Ca I gas then the models of Lodders (2002) imply a
brightness temperature 1820-1850 K for an L1 in $K$-band. The Al
chemistry is more complicated, but a temperature in the range
1950-2000 K would appear possible at $J$. In any case, the
presence of Ca I in the $K$-band as late as L1, combined with the
absence of Al I in the $J$-band spectrum of an L0 should provide a
useful constraint on the temperature at the M/L boundary, with the
caveat that different temperature layers are being probed at these
two wavelengths.

L and T dwarfs are not black bodies, and brightness temperatures
derived from spectra can range over 1000 K in the near-infrared
due to the very large variations in opacity with wavelength
(Saumon et al. 2000). In the absorption bands of H$_{2}$O, one
probes only the cooler upper levels of the atmosphere, whereas in
the continuum near the pressure broadened K I lines the emergent
flux comes from much deeper layers. The condensation chemistry
temperatures discussed above are significantly lower ($\sim$500K)
than those derived by empirical estimates of T$_{eff}$ (from
luminosity and radius) for objects with known parallax (B02).
Using that method, B02 provide a correlation between temperature
and spectral type for L dwarfs, and find that extrapolation of the
roughly linear relation between spectral type and effective
temperature for L dwarfs significantly underestimates the
temperatures of T dwarfs. The almost linear behavior of the
strength of the H$_{2}$O (and CH$_{4}$) bands with spectral type
presented here does not necessarily imply a continuous linear
variation of T$_{eff}$ with spectral type. Dust formation and
settling has a significant effect. Clearing of dust as the
photospheric temperature falls, enables the column density of
water (and methane) to increase steadily. Also, conversion of CO
to \ch4 produces additional \h2o to strengthen the bands. B02
speculate that only a narrow temperature range separates L8-T5
objects.

There are also objects which stand out as exceptions to the trends
we have identified. One of these is the L6.5 object 2MASS 2244+20.
Figure 24 compares the 1.1-2.3 \um spectra of this object and two
other sources classified as L6 and L7. All three spectra have been
normalized at 1.270 \micron~ in the $J$-band. Not only is the flux
from 2MASS 2244+20 dramatically enhanced at $H$ and $K$ relative
to the levels expected for an L6.5, but the \h2o is weaker and the
CO bands are unusually strong. A weighted average of the \h2o
indexes gives L5. In addition, Table 6 and Figs. 14-15 show that
2MASS 2244+20 has the smallest K I equivalent widths of any L
dwarf (2-3 \AA). This result may imply a lower surface gravity or
perhaps a lower metallicity. Finally, the $H$-band is clearly
peaked at $\sim$1.70 \micron~ unlike any other L or T dwarf. This
L dwarf, which has extremely red IR colors ($J-K_{s}=2.48\pm
0.15$; $H-K_{s}=1.04\pm 0.10$), clearly has unusual IR spectral
features. In contrast, the far-red optical spectrum of 2MASS
2244+20 obtained by Kirkpatrick using the LRIS spectrograph on
Keck appears normal when compared to other L dwarfs (see K99 for
methods). Figure 25 shows the LRIS spectrum compared to that of
2MASS 0103+19 and D0205-11. Overall, the infrared characteristics
of 2MASS 2244+20 may be indicative of unusually strong veiling
attributable to dust clouds or perhaps an unusual metallicity.

Major changes in opacity caused by the formation and subsequent
clearing of dust clouds would also explain the behavior of the
flux-calibrated spectra in Figs. 3 \& 10. The $J$-band flux
becomes more and more depressed relative to $H$ and $K$ (hence the
colors redden) as the spectral type changes from M to late L.
Around L8 the flux peaks in $F_{\lambda}$ are broadly level in $J,
H$ and $K$. This situation is reversed at the L/T boundary where
the $J$-band flux becomes progressively stronger again as the T
spectral subclass becomes later; the peak fluxes in each band are
always such that $J>H>K$. In addition, the equivalent widths of
the K I lines increase again in the early T's after a decline
through the late L types, consistent with seeing deeper into a
clearer atmosphere. These trends are verified by the parallax data
of Dahn et al.\ (2002), which show a brightening at $J$ from L8 to
T5. Burgasser et al.\ (2002a) have attributed this brightening to
the rapid dispersal and condensation of clouds across the L/T
transition.

The six molecular band indexes used here perform remarkably well
over a large spectral range. Two of the \h2o indexes and one \ch4
index saturate at the current T8, while the remaining indexes
appear to have some more range. It is very likely however, that
some of these ratios will turn around and increase. That is, they
will no longer be monotonic over a wider range of later types.
Such behavior would arise even though the absorber in question
saturates the ``on-band" portion of the index, because additional
absorbers may begin to suppress the ``off-band" part of the index.
For example, in the $K$-band, where \ch4 appears to saturate at
T8, it is possible that CIA by H$_{2}$ will continue to suppress
the ``off-band" pseudo-continuum peak, eventually flattening the
entire band. Termination of the T sequence may be more influenced
by \h2o than \ch4, since the formation of water clouds near
$\sim$500 K may have a significant effect on the spectra.

\section{CONCLUSIONS}
We have obtained and analyzed a sample of 53 $J$-band spectra and
25 1-2.3 \micron~ spectra of M, L and T dwarfs with high
signal-to-noise, identical instrumentation, and a uniform data
reduction process. A relatively high spectral resolution of R
$\sim$2,000 was employed in our NIRSPEC survey, yielding a library
of spectra more useful than lower resolution data for constraining
model atmospheres.

Nine near-infrared molecular bands (4 \h2o, 2\ch4, 2FeH and 1 CO),
together with several neutral atomic species (K, Na, Fe, Al, Ca),
appear to provide good diagnostics for a self-consistent, pure
infrared, spectral classification scheme for both L and T dwarfs
in most cases. Lines of Al I in the $J$-band disappear at the M/L
boundary. The consistent appearance of \ch4 in both the $H$ and
$K$ bands defines the L/T boundary, but evidence of \ch4 can be
seen as early as L7 in some objects. Flux ratio indexes for the
\h2o and \ch4 features are monotonic and almost linear with
spectral class. Consistency with the optically-derived spectral
types suggests that the primary underlying physical quantity
controlling the appearance of the spectra is effective
temperature. The K I lines are gravity sensitive. Flux-calibrated
spectra for 25 sources, 16 of which provide overlapping wavelength
coverage with far red optical spectra, show that the peak flux
emerges in the $J$-band for early L dwarfs and for T dwarfs, but
that late L dwarfs have significantly suppressed $J$-band flux,
consistent with redder colors. One source, 2MASS 2244+20,
classified optically as L6.5 has discordant molecular band flux
ratios and exhibits a spectrum that is more depressed in the
$J$-band than any other L dwarf. This behavior is probably due to
dust extinction, but the effects of metallicity also need to be
considered.

The spectral classification of main-sequence stars, as originally
codified in the Henry Draper catalog (Pickering 1890; Cannon and
Pickering 1901) and refined into the MK system (Morgan, Keenan,
and Kellman 1943), requires the establishment of spectral
standards. Once the standards are established, the response of a
given spectrograph at a given site can be calibrated, and objects
of unknown types compared to the standards, as observed by that
instrument at that site. K99 established standards for L dwarfs
using far red optical diagnostics. B02 provided a set of standards
for T dwarfs based on infrared properties. By including many of
these standards in the BDSS, we have demonstrated that a
self-consistent pure NIR classification scheme should be possible.

In subsequent papers we will report on spectra with ten times
higher resolution (McLean et al.\ 2003b; in prep.), discuss the
establishment of a near-infrared L classification scheme and
compare data to models. In addition, we will report on the results
of programs to (1) observe the morphology and spectral
characteristics of a subsample of bright brown dwarfs at
R$\sim$20,000 (McLean et al.\ 2003b; in prep.), (2) develop
indicators of surface gravity (McGovern et al. 2003; in prep.),
and (3) use high resolution spectroscopy to search for radial
velocity variations indicative of spectroscopic companions (Prato
et al. 2003; in prep.). The BDSS data will be made available on
request and through an archive web site currently under
development
(\url{http://www.astro.ucla.edu/$\sim$mclean/BDSSarchive}).

\acknowledgments The authors wish to thank the staff of the Keck
Observatory for their outstanding support, in particular, our
Observing Assistants Joel Aycock, Ron Quick, Gary Puniwai,Cynthia
Wilburn, Chuck Sorenson, Gabriel Saurage and Terry Stickel, and
our Instrument Scientists David Sprayberry, Grant Hill, Randy
Campbell, Bob Goodrich and Paola Amico. The authors are grateful
to those of Hawaiian ancestry on whose sacred mountain we are
privileged to be guests. ISM acknowledges the staff of the UCLA
Infrared Lab, and colleagues James Graham (UCB), Eric Becklin
(UCLA) and James Larkin (UCLA) for support throughout the
development of the NIRSPEC instrument. We are also grateful to
Adam Burrows, Mark Marley, Didier Saumon and Katharina Lodders for
useful discussions on model atmospheres. AJB acknowledges support
by NASA through Hubble Fellowship grant HST-HF-01137.01 awarded by
the Space Telescope Science Institute, which is operated by the
Association of universities for research in Astronomy, Inc., for
NASA, under contract NAS 5-26555. This publication makes use of
data from the Two Micron All Sky Survey, which is a joint project
of the University of Massachusetts and the Infrared Processing and
Analysis Center, funded by the National Aeronautics and Space
Administration and the National Science Foundation. Work by SSK
was supported by the Astrophysical Research Center for the
Structure and Evolution of the Cosmos (ARCSEC) of Korea Science
and Engineering Foundation through the Science Research Center
(SRC) program.

\clearpage

\begin{figure}[!htp]
\epsscale{0.95} \plotone{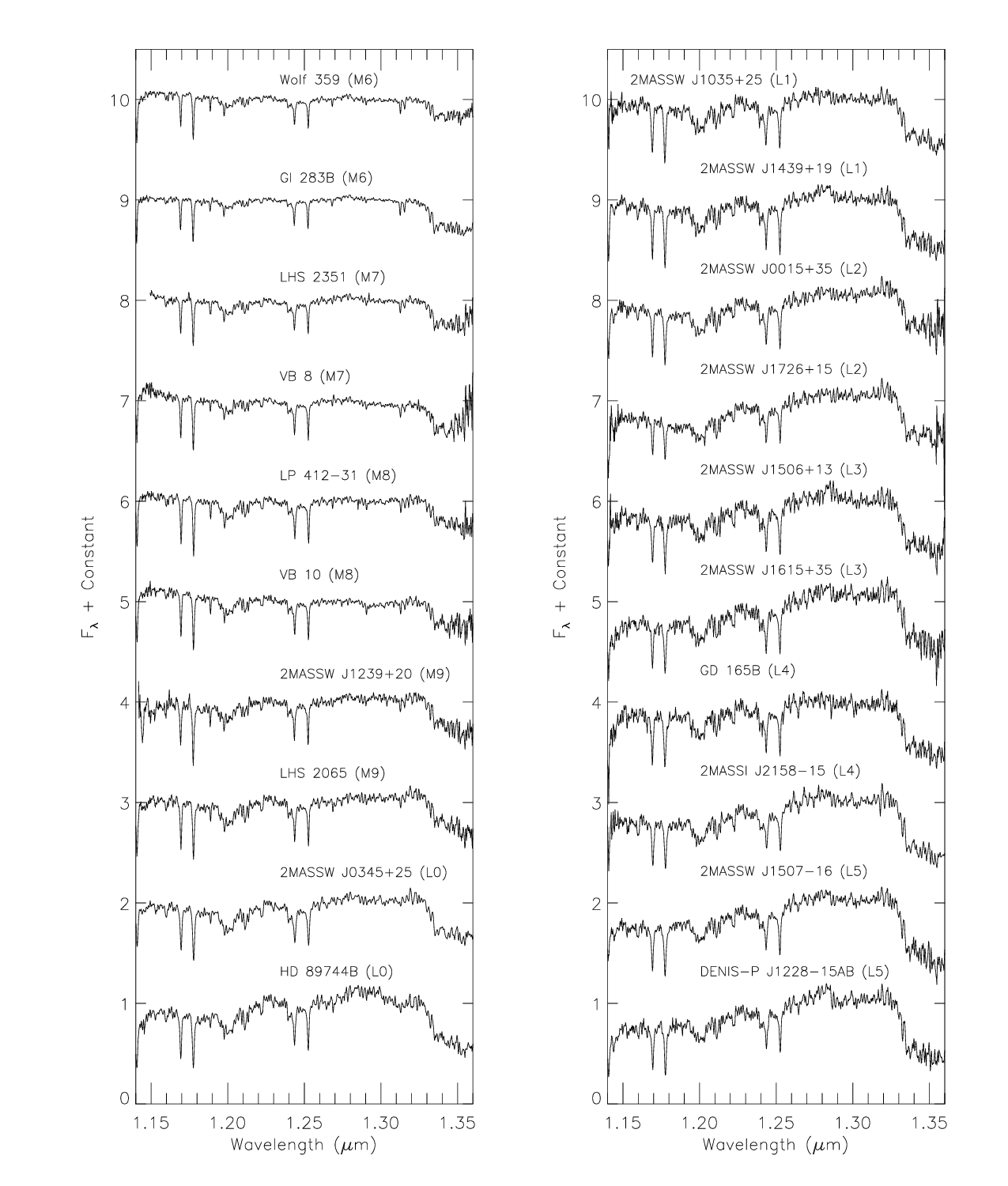} \caption{NIRSPEC $J$- band
sequence for spectral types M6-L5. L dwarf classifications are
from K99. Two objects are represented for each spectral type bin.
Spectra are normalized to 1.0 at 1.265 \um and offset by integers
on the flux axis for clarity. Extended tick marks on the vertical
axis denote zero flux levels for each spectrum. The
signal-to-noise ratio is typically $>$ 30 and essentially all of
the fine spectral structure is real and repeatable.}
\end{figure}
\epsscale{1}

\begin{figure}[!htp]
\epsscale{0.95} \plotone{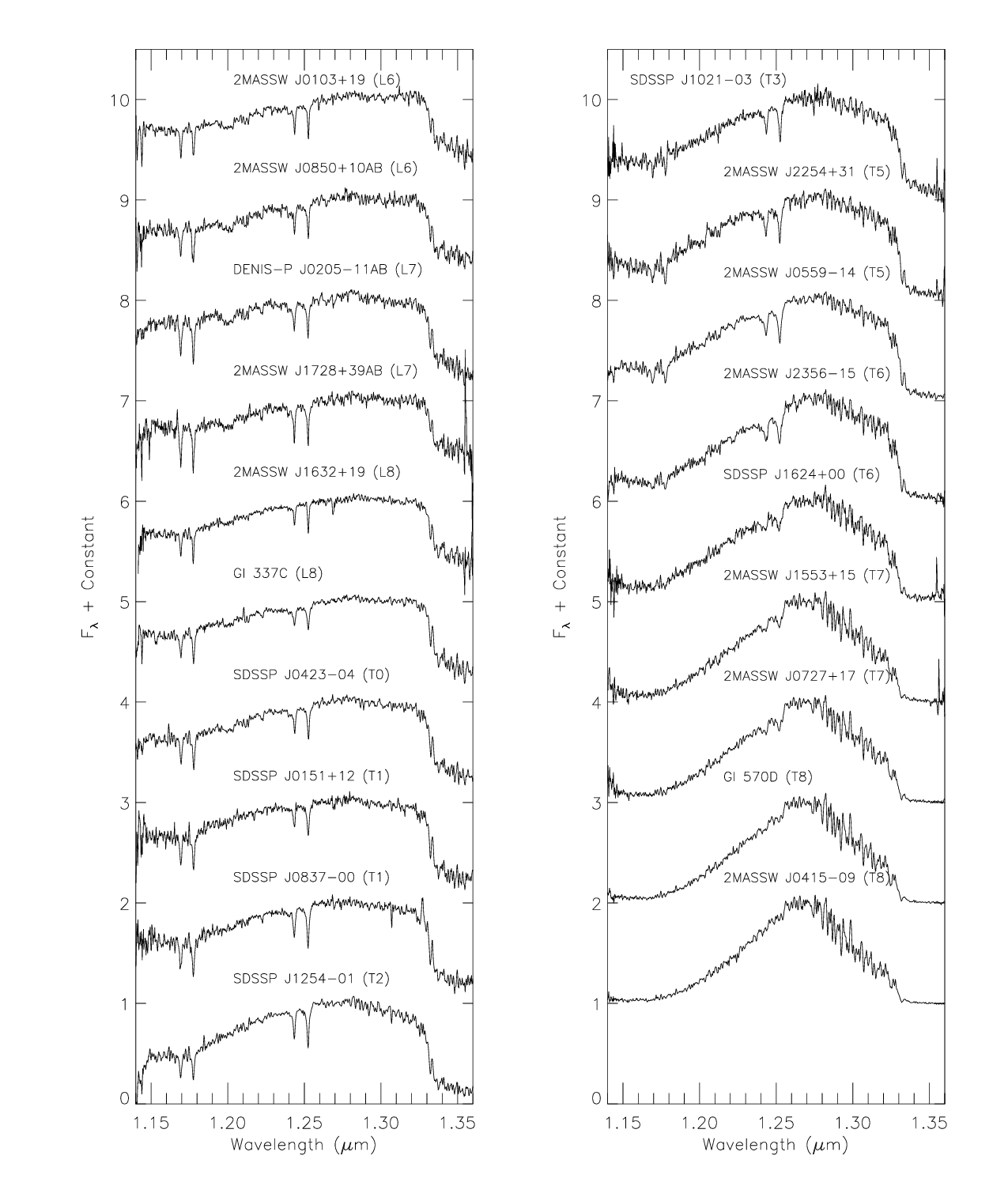} \caption{NIRSPEC $J$- band
sequence for spectral types L6-T8.  Two objects are represented
for each spectral type bin where possible. Classifications are
from K99 for L dwarfs and B02 for T dwarfs. There is no defined L9
class. Only one T0, T2, and T3 are known to date. No examples of
T4 have yet been discovered. As in Figure 1, spectra are
normalized to 1.0 at 1.265 \um and offset by integers. Extended
tick marks on the vertical axis denote zero flux levels.
Essentially all of the fine spectral structure is real and
repeatable.}
\end{figure}
\epsscale{1}

\begin{figure}[!htp]
\epsscale{0.9} \plotone{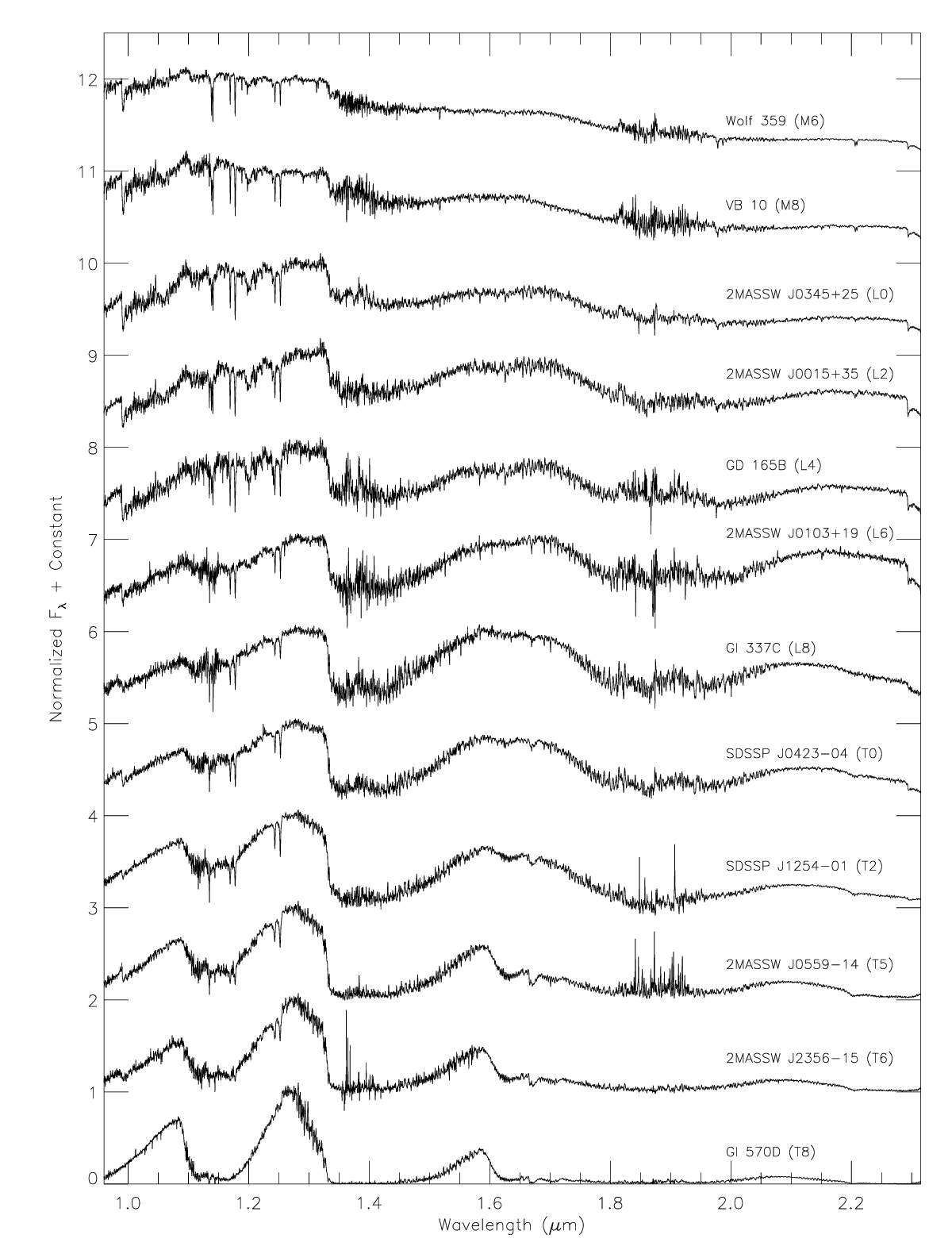} \caption{NIRSPEC spectral sequence
for even-typed subclasses from M6 - T8 covering wavelengths 0.96 -
2.315 \micron. No T4 dwarf is known to exist to date; so in place
of a T4 the spectra of 2MASSW J0559-14 (T5) is shown. Spectra are
flux-calibrated, then normalized to 1.0 at 1.270 \um and offset by
integers for display. Flux normalization constants are given in
Table 3. Extended tick marks on the vertical axis denote zero flux
levels. Noisy regions at 1.15, 1.40 and 1.90 \um are the result of
low atmospheric transmission.}
\end{figure}
\epsscale{1}

\begin{figure}[!htp]
\epsscale{0.9}
\plotone{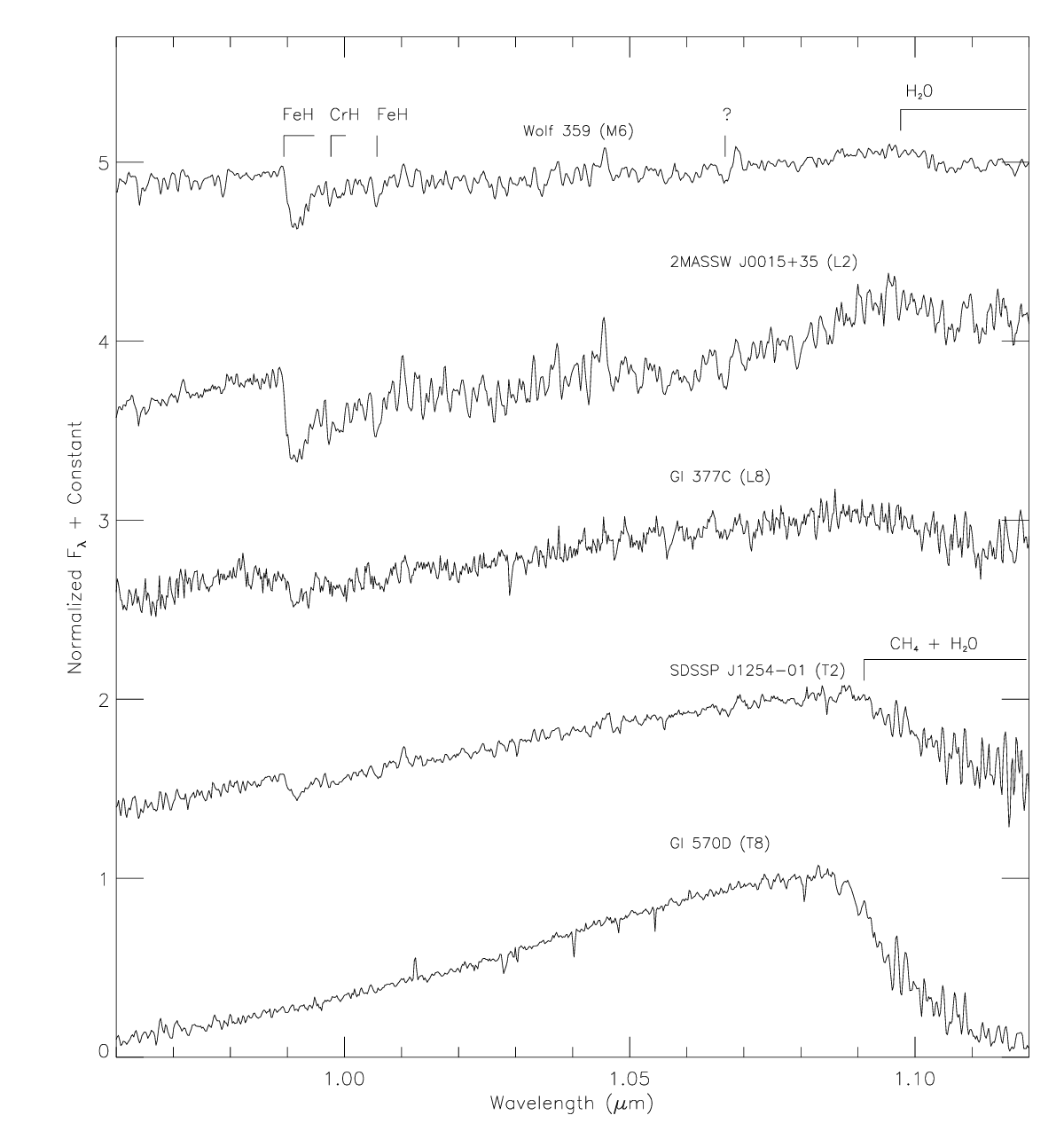} \caption{$z$-band spectra of
Wolf 359 (M6), 2MASSW 0015+35 (L2), Gl 337C (L8), SDSS 1254-01
(T2), and Gl 570D (T8). Spectra are normalized at 1.08 \um and
offset vertically by integers. Extended tick marks on the vertical
axis denote zero flux levels for each spectrum.  Prominent FeH,
CrH, \h2o and \ch4 molecular features are identified.}
\end{figure}
\epsscale{1}

\begin{figure}[!htp]
\epsscale{0.9} \plotone{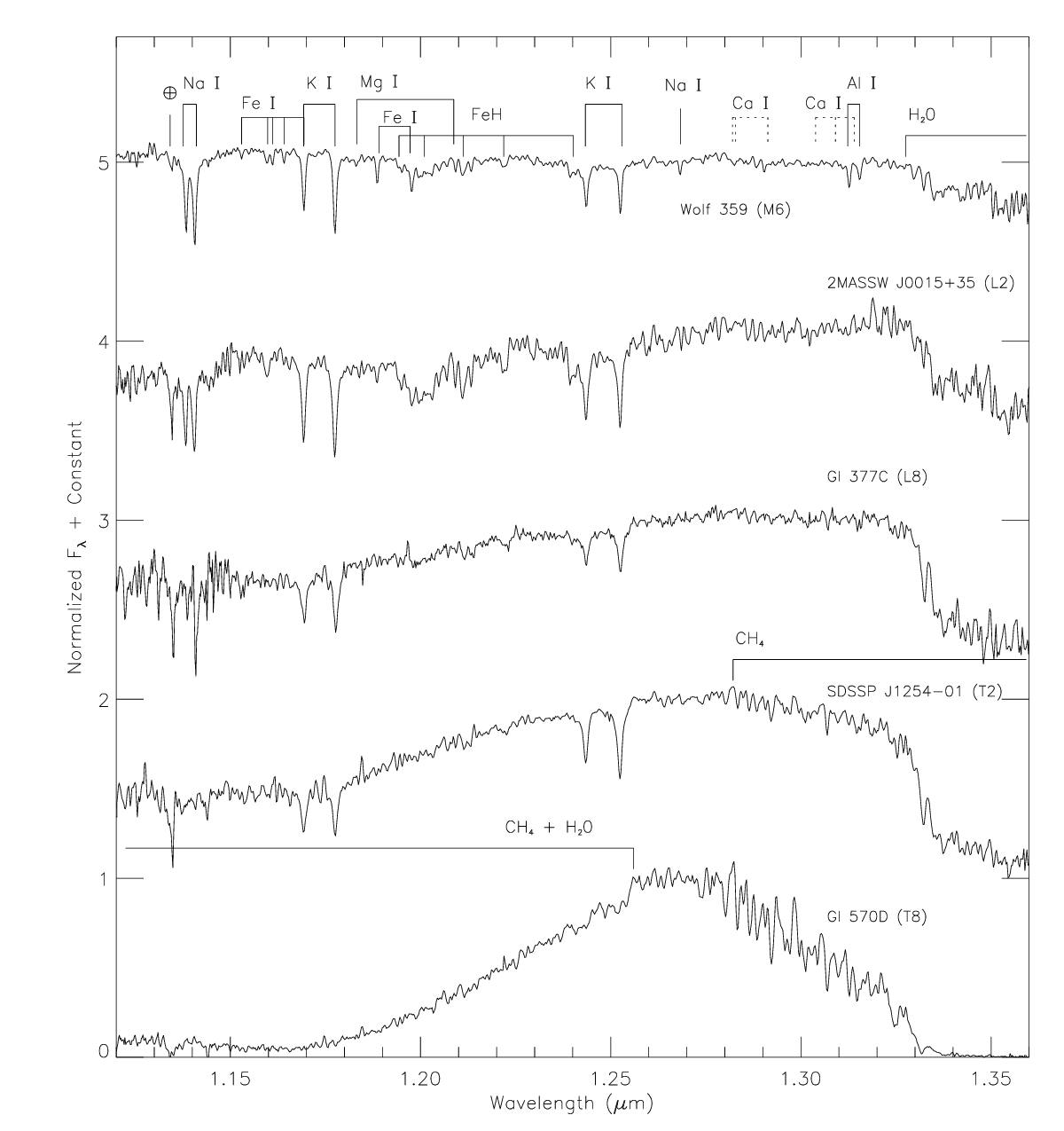} \caption{$J$-band spectra of Wolf
359 (M6), 2MASSW 0015+35 (L2), Gl 337C (L8), SDSS 1254-01 (T2),
and Gl 570D (T8). Spectra are normalized to 1.0 at 1.265 \um and
offset vertically by integers. Extended tick marks on the vertical
axis denote zero flux levels.  Prominent \h2o, \ch4, and FeH
molecular features, and atomic lines of Na I, K I, Fe I, Mg I and
Al I are identified.}
\end{figure}
\epsscale{1}

\begin{figure}[!htp]
\epsscale{0.9} \plotone{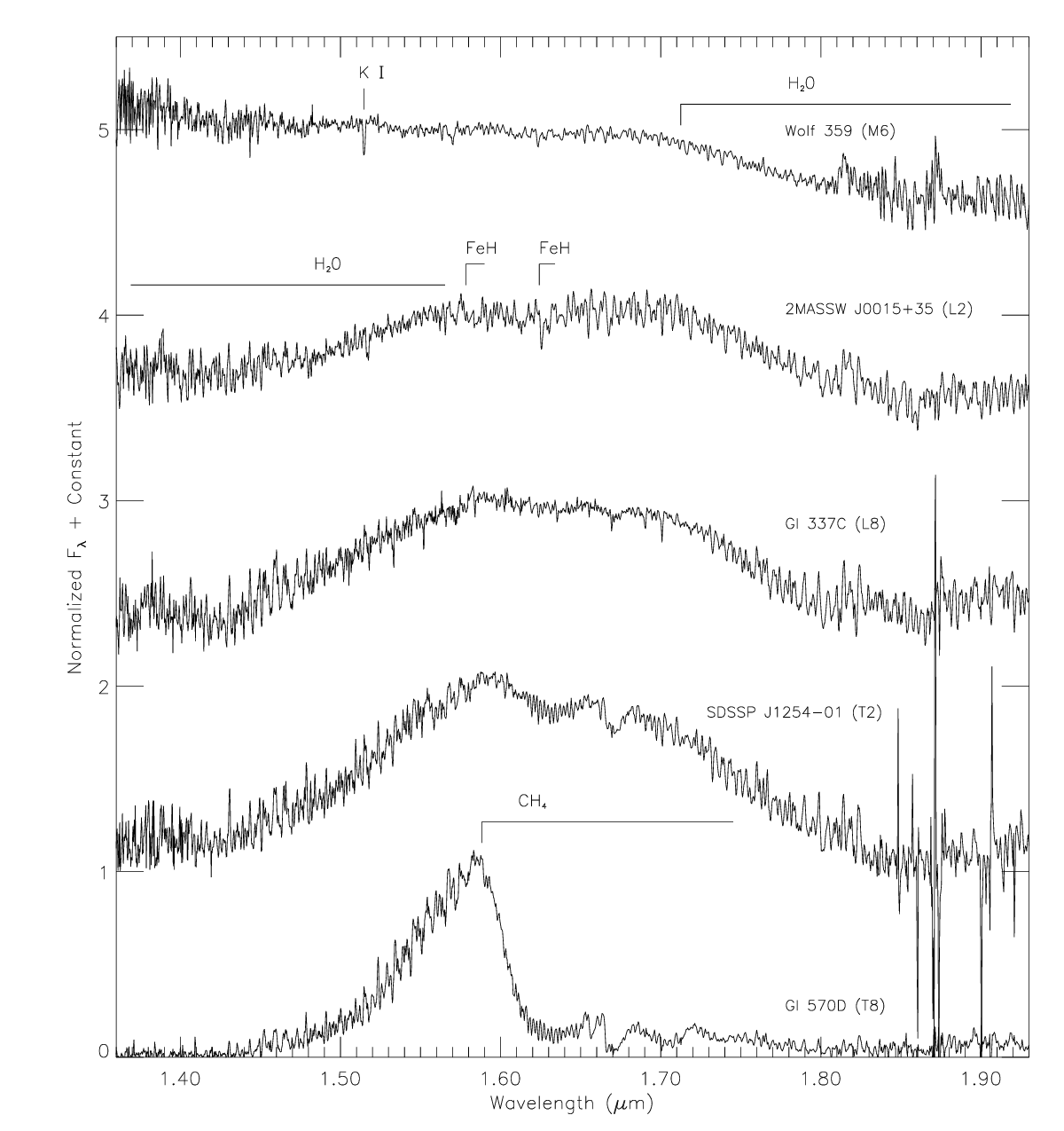} \caption{$H$-Band spectra of Wolf
359 (M6), 2MASSW 0015+35 (L2), Gl 337C (L8), SDSS 1254-01 (T2),
and Gl 570D (T8).  Spectra are normalized to 1.0 at 1.583 \um and
offset vertically by integers.  Extended tick marks on the
vertical axis denote zero flux levels.  Molecular features of
\h2o, \ch4, and FeH are identified together with a line from K I.
Essentially all of the fine structure is real and repeatable.}
\end{figure}
\epsscale{1}

\begin{figure}[!htp]
\epsscale{0.9}
\plotone{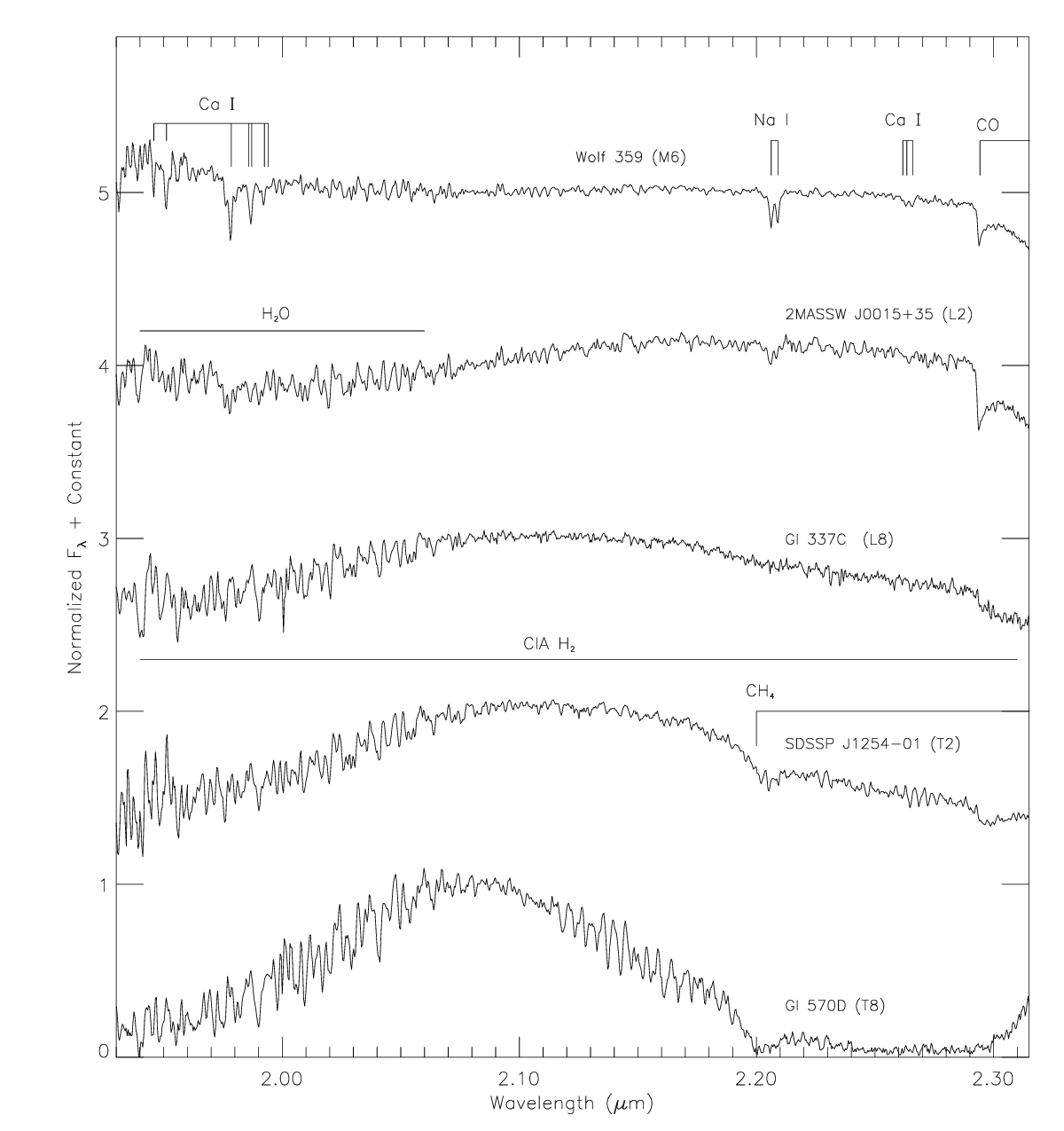} \caption{$K$-Band spectra of
Wolf 359 (M6), 2MASSW 0015+35 (L2), Gl 337C (L8), SDSS 1254-01
(T2), and Gl 570D (T8).  Spectra are normalized to 1.0 at 2.080
\um and offset vertically by integers.  Extended tick marks on the
vertical axis denote zero flux levels.  \h2o, CO, and \ch4
molecular features are marked, together with Ca I and Na I atomic
lines. Collision induced absorption by H$_2$ covers essentially
all of $K$-band.}
\end{figure}
\epsscale{1}

\begin{figure}[!htp]
\epsscale{0.9}
\plotone{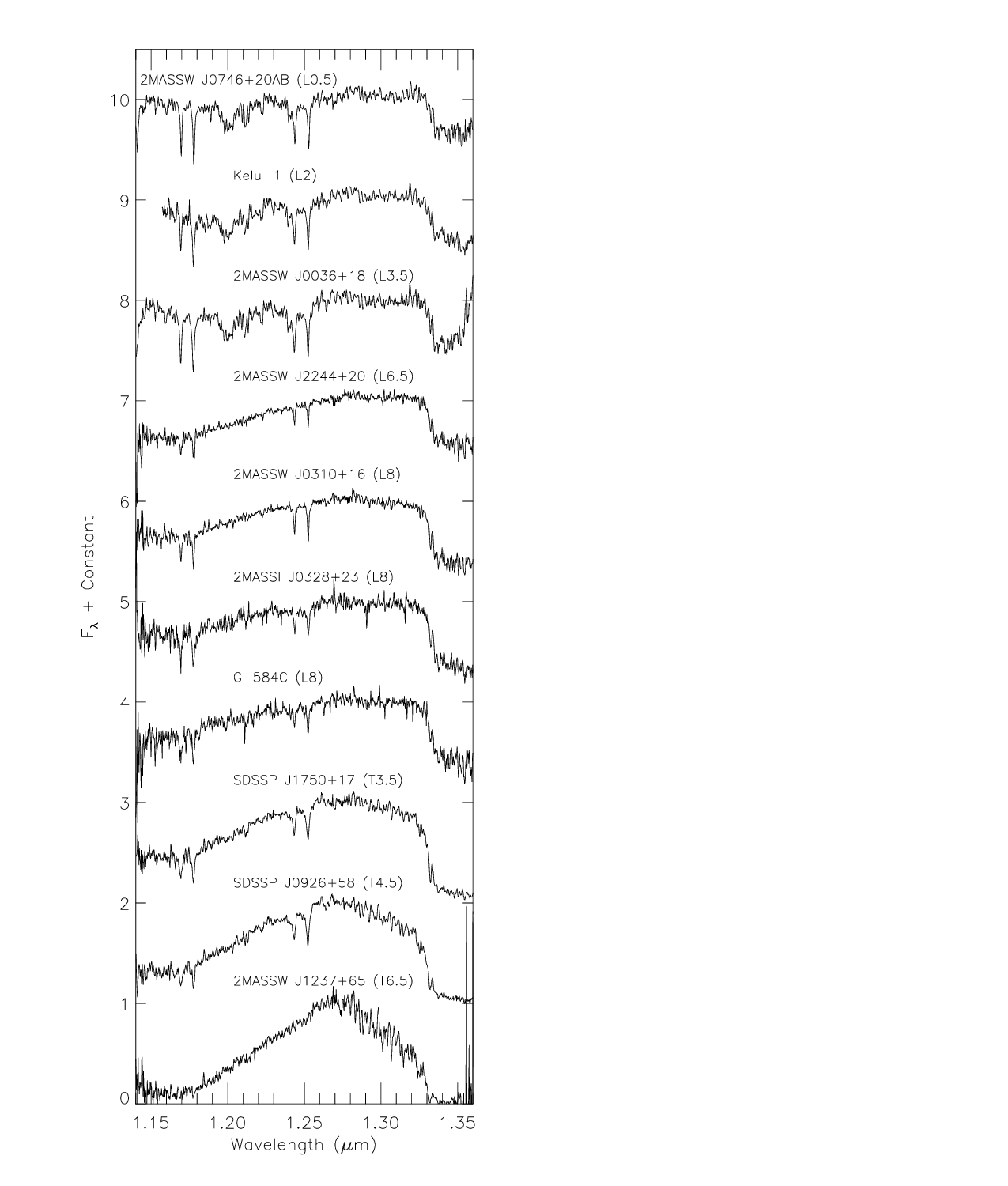} \caption{Additional
$J$-band spectra. Objects include all half types from the survey
(6), additional L8 dwarfs and a L2 dwarf (Kelu-1). Objects are
plotted on the same scale as the main $J$-band figures for ease of
comparison. Spectra are normalized to 1.0 at 1.265 \um and offset
by integers.  Extended tick marks on the vertical axis denote zero
flux levels.}
\end{figure}
\epsscale{1}

\begin{figure}[!htp]
\epsscale{0.9}
\plotone{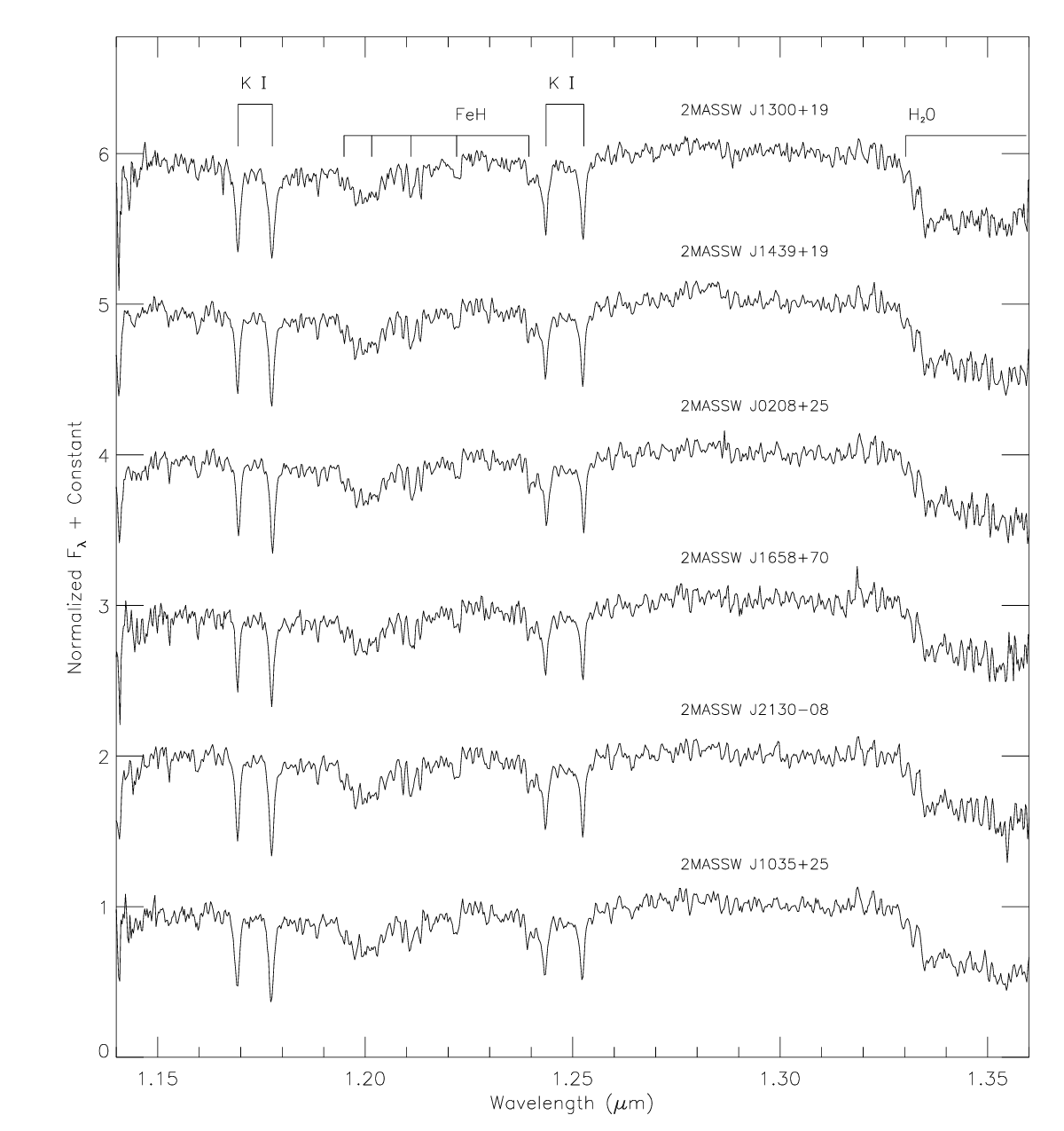} \caption{Spectra of six L1
dwarfs are plotted to investigate diversity in the near-infrared
among objects of equivalent optically-based classification.
Spectra are normalized to 1.0 at 1.265 \um and offset by integers
as in previous figures. Extended tick marks on the vertical axis
denote zero flux levels.}
\end{figure}
\epsscale{1}

\begin{figure}[!htp]
\epsscale{0.9} \plotone{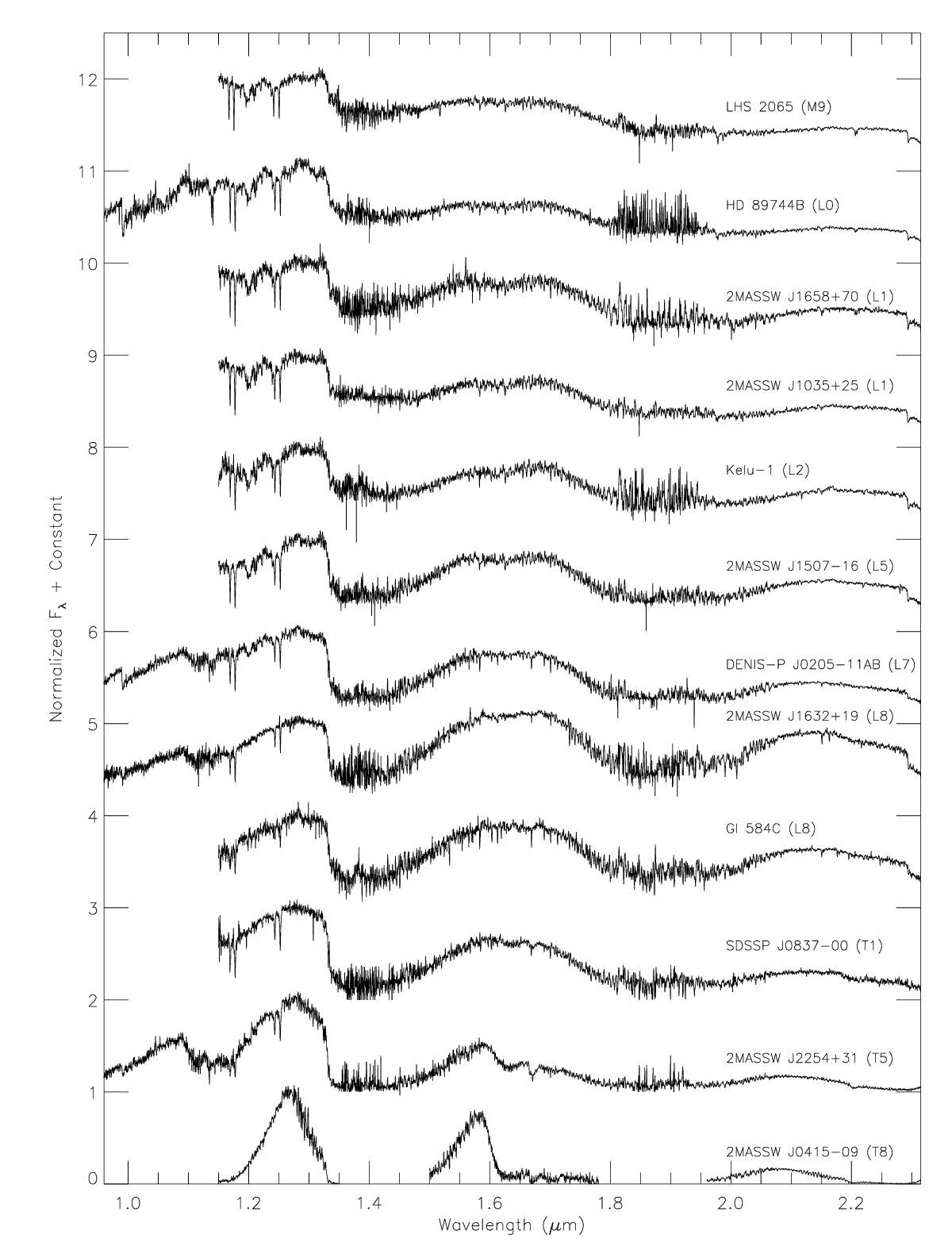} \caption{$JHK$ spectra
(1.14-2.315 \micron) of an additional 12 objects in the BDSS in
the spectral range M9-T8. Spectra are plotted on the same scale as
Figure 3 for comparison. Where available, additional $z$-band data
are plotted to extend the spectral coverage to 0.96 \micron.
Spectra are normalized to 1.0 at 1.270 \um and offset by integers.
Flux normalization values are given in Table 3. Extended tick
marks on the vertical axis denote zero flux levels.}
\end{figure}
\epsscale{1}

\begin{figure}[!htp]
\epsscale{0.9} \plotone{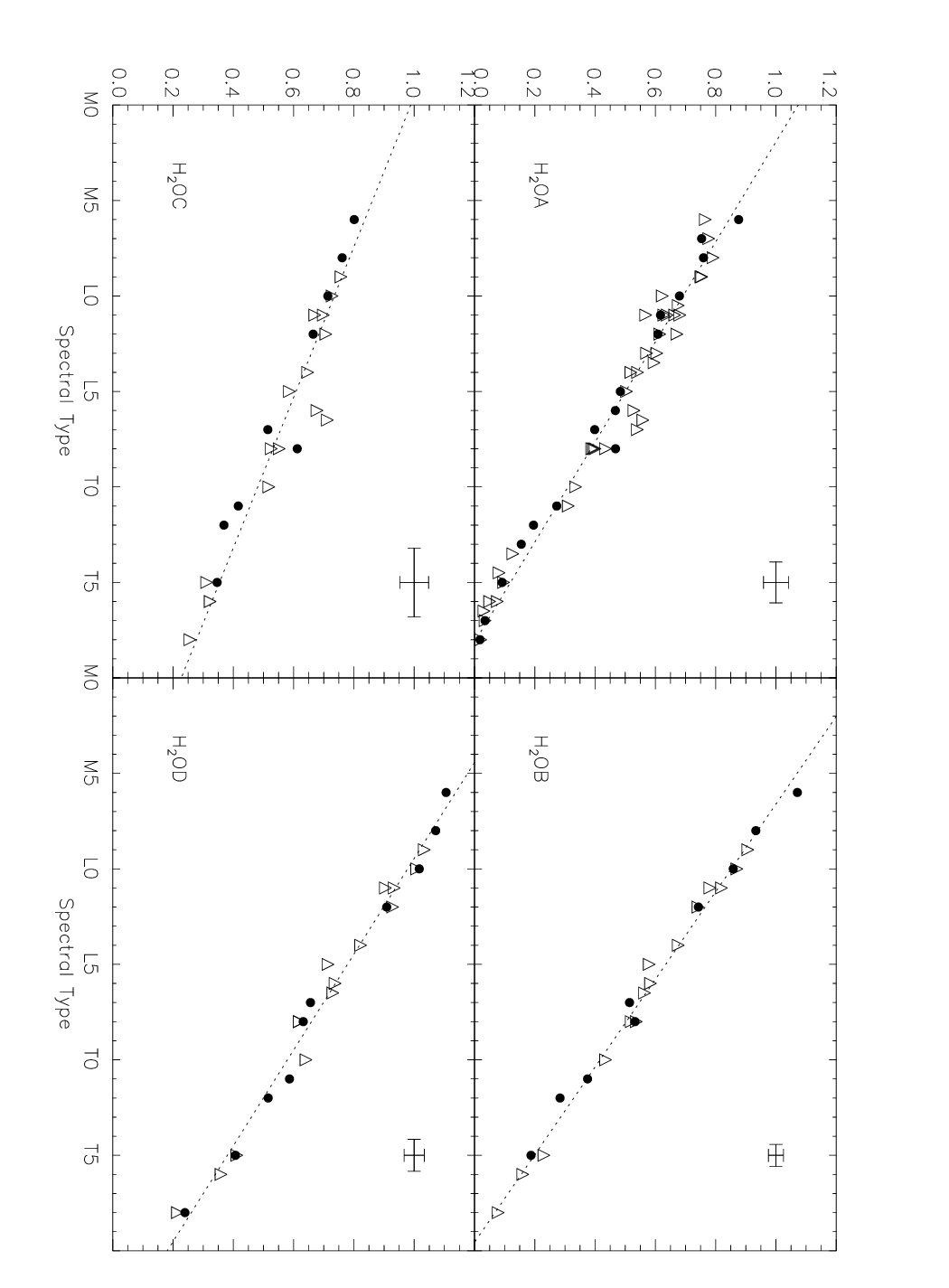} \caption{Four flux ratios based
on the \h2o bands at (A) 1.34 \micron, (B) 1.456 \micron, (C)
1.788 \micron, and (D) 1.964 \micron~ are plotted as a function of
published spectral type. Solid symbols are spectral classification
standards from K99 and B02 (see text). The major outlier at L6.5
in the \h2oC plot is 2MASS 2244+20. See Table 5 for a listing of
the objects corresponding to the plotted points. Dashed lines show
best fitting straight line and error bars in each corner are $\pm
1 \sigma$ errors derived from the residuals to the linear fit.}
\end{figure}
\epsscale{1}

\begin{figure}[!htp]
\epsscale{0.8} \plotone{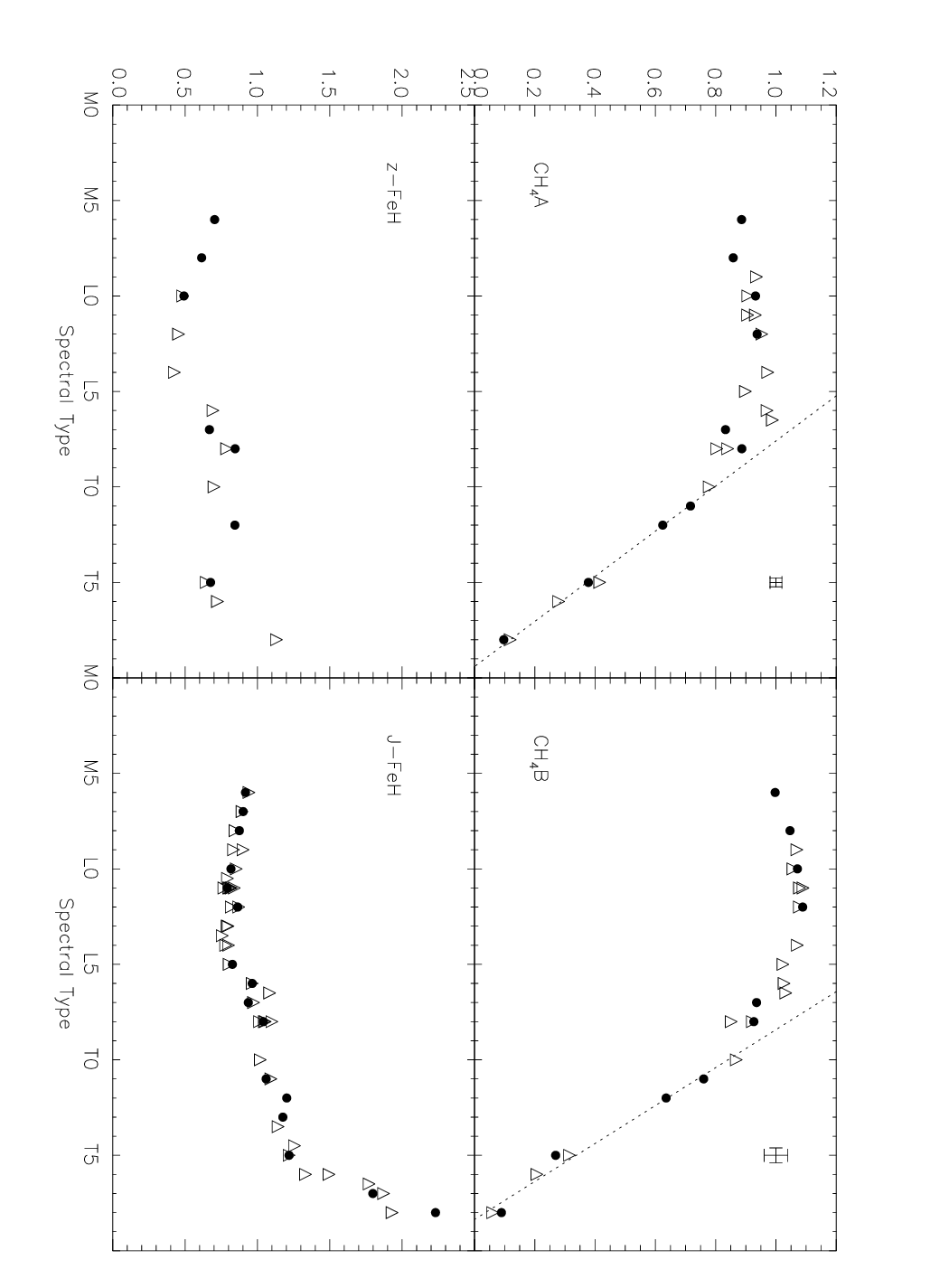} \caption{Upper panels: the flux
ratio for (A) the 1.730 \micron~ \ch4 band and (B) the 2.200
\micron~ \ch4 band as a function of published spectral type. Lower
panels: The flux ratio for the 0.992 \micron~ FeH band ($z$-FeH)
and the flux ratio for the 1.200 \micron~ FeH band ($J$-FeH) as a
function of published spectral type. Solid symbols are spectral
classification standards from K99 and B02 (see text). Best fitting
straight lines are shown where appropriate. Error bars in the
corners are $\pm 1 \sigma$ for the best fitting line.}
\end{figure}
\epsscale{1}

\clearpage

\begin{figure}[!htp]
\epsscale{0.9} \plotone{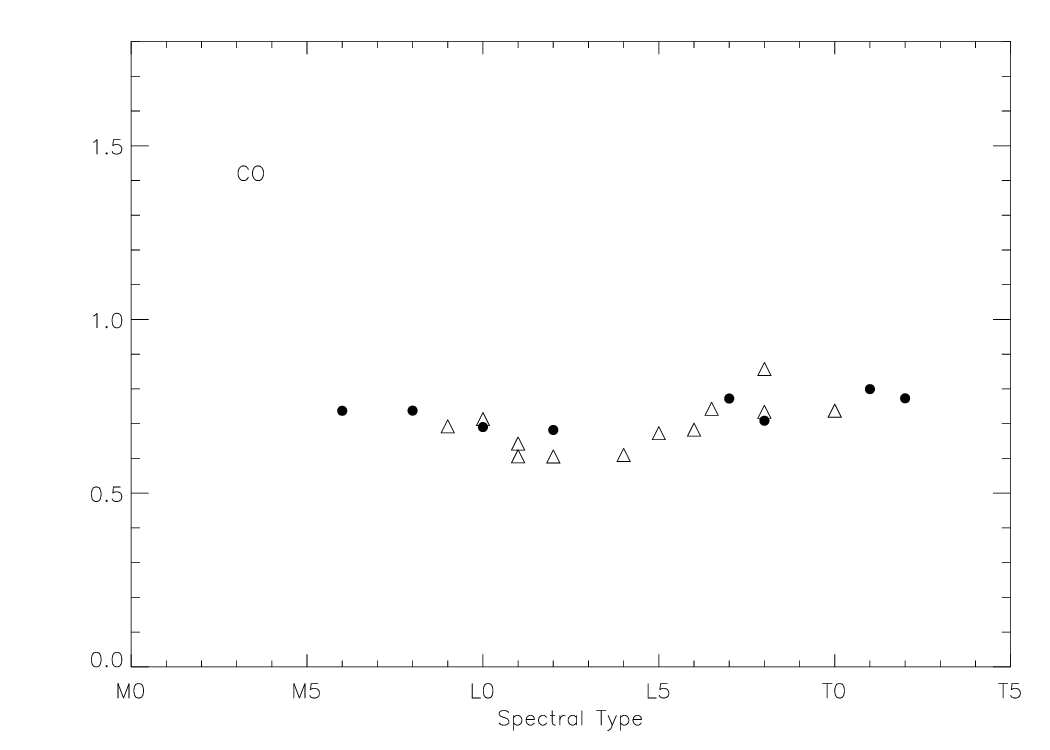} \caption{The flux ratio for the
2.295 \micron~ (v=2-0) CO band as a function of published spectral
type. Solid symbols are spectral classification standards from K99
and B02. The ratio is contaminated by the onset of \ch4 absorption
in T dwarfs at the continuum wavelength and not useful beyond T3.}
\end{figure}
\epsscale{1}

\begin{figure}[!htp]
\epsscale{0.9} \plotone{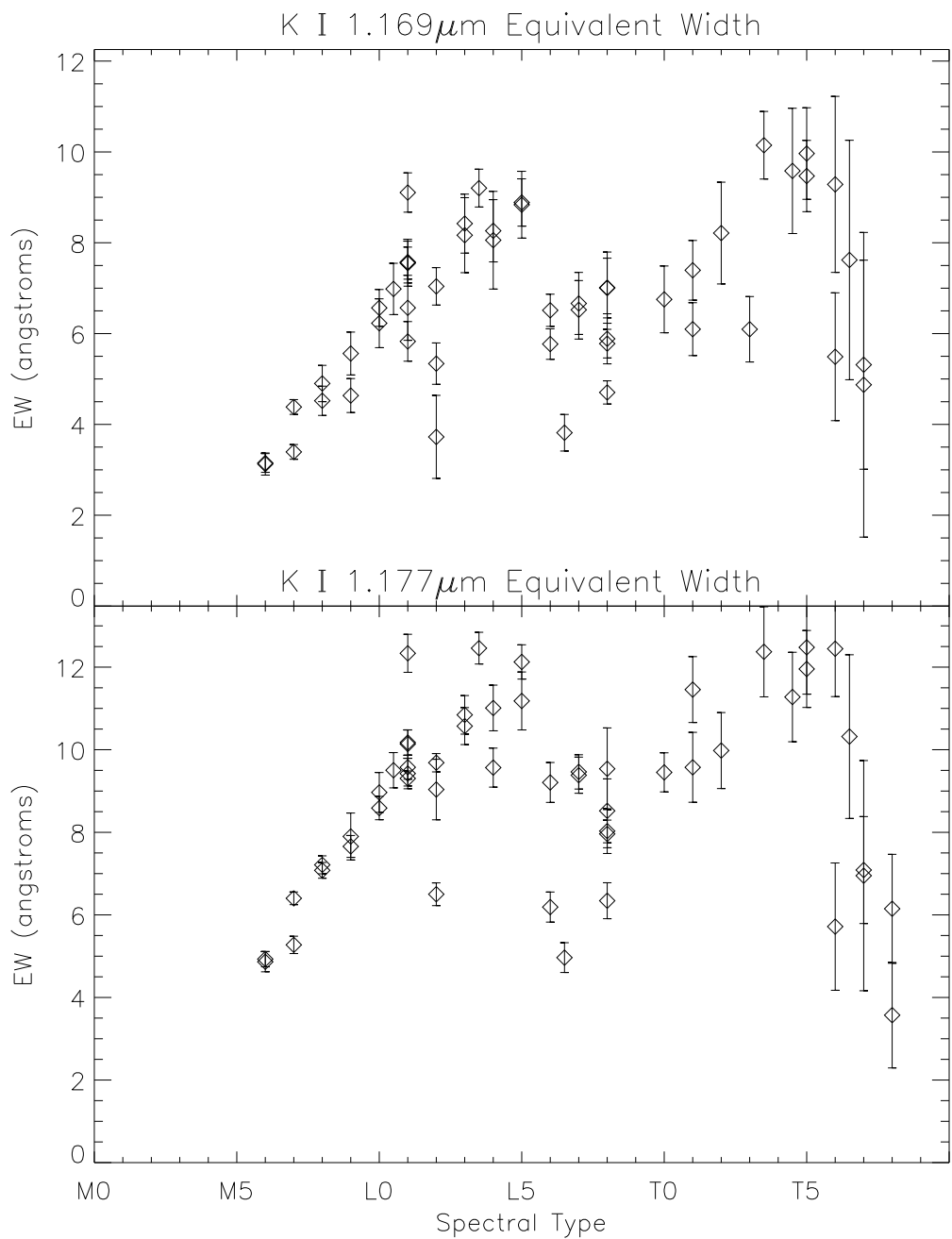} \caption{Equivalent width of the
1.169 \um and 1.177 \um lines of K I as a function of published
spectral type in the K99 and B02 systems. The L2 dwarf with the
smallest EW in the lower panel is 2MASS1726+15, and the L6.5 with
a weak 1.177 $\mu$m line is 2MASS2244+20. Kelu-1 (L2) is also
lower than expected.}
\end{figure}
\epsscale{1}

\begin{figure}[!htp]
\epsscale{0.9} \plotone{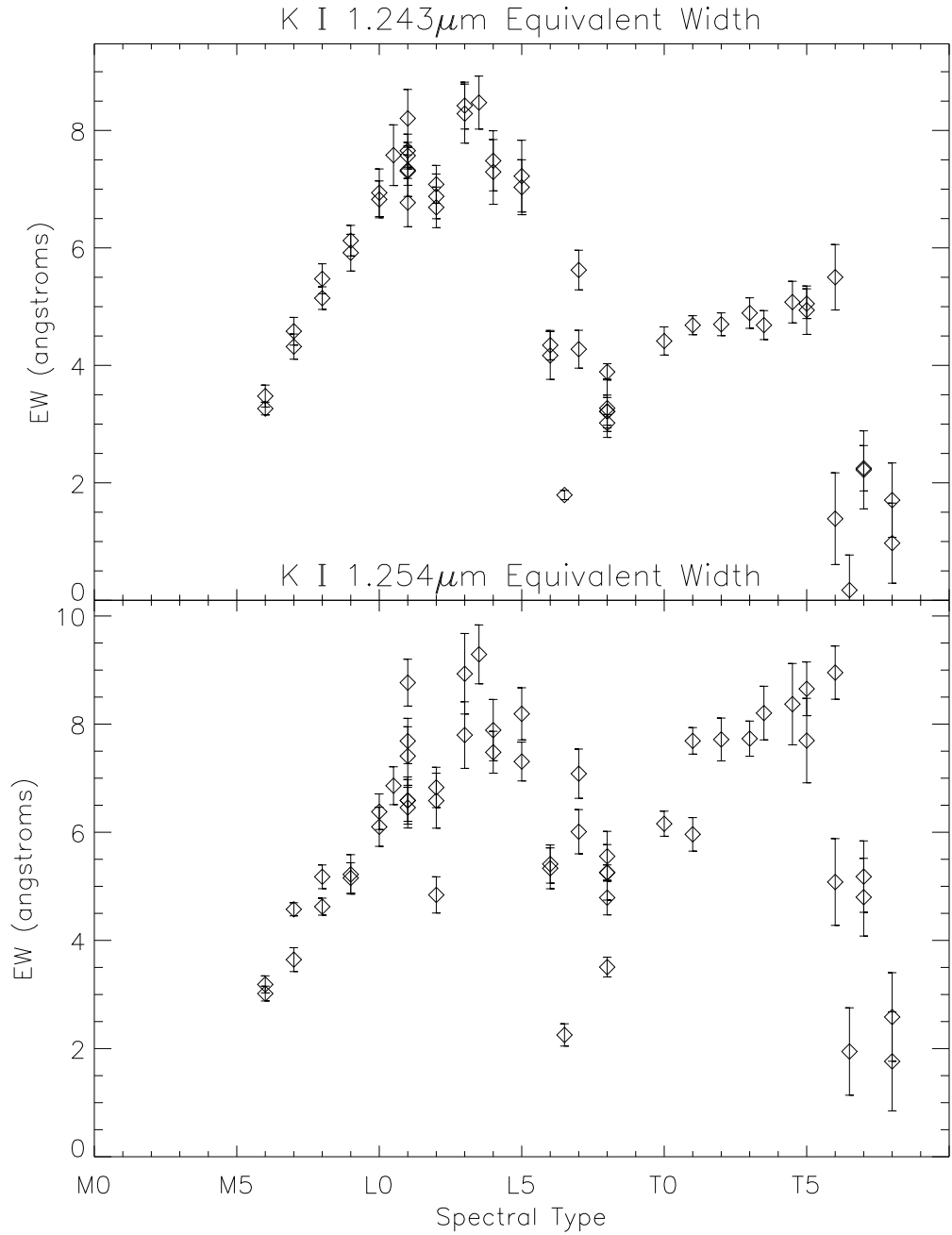} \caption{Equivalent width of the
1.243 \um and 1.252 \um lines of K I as a function of published
spectral type in the K99 and B02 systems. Contamination by FeH has
been minimized for the 1.243 \um line. Again, the L2 and L6.5 with
weak lines are 2MASS1726+15 and 2MASS2244+20 respectively.}
\end{figure}
\epsscale{1}

\begin{figure}[!htp]
\epsscale{0.9} \plotone{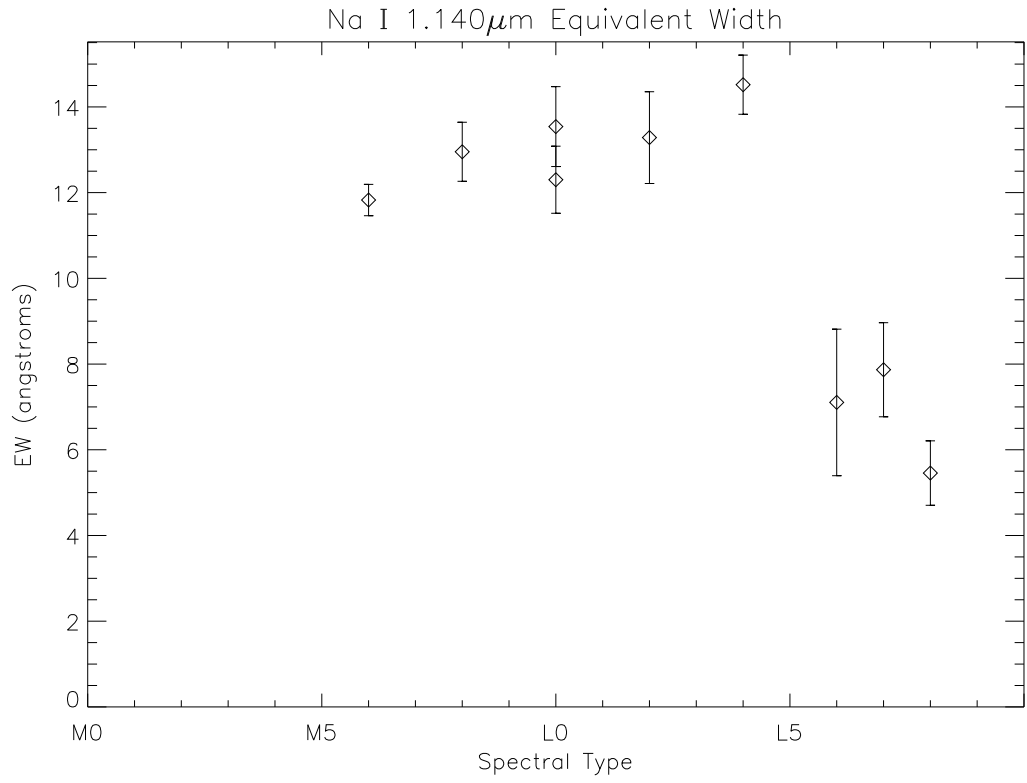} \caption{Equivalent width of the
1.14 \um Na I doublet as a function of published spectral type in
the K99 and B02 systems. Both lines are combined.}
\end{figure}
\epsscale{1}

\begin{figure}[!htp]
\epsscale{0.9} \plotone{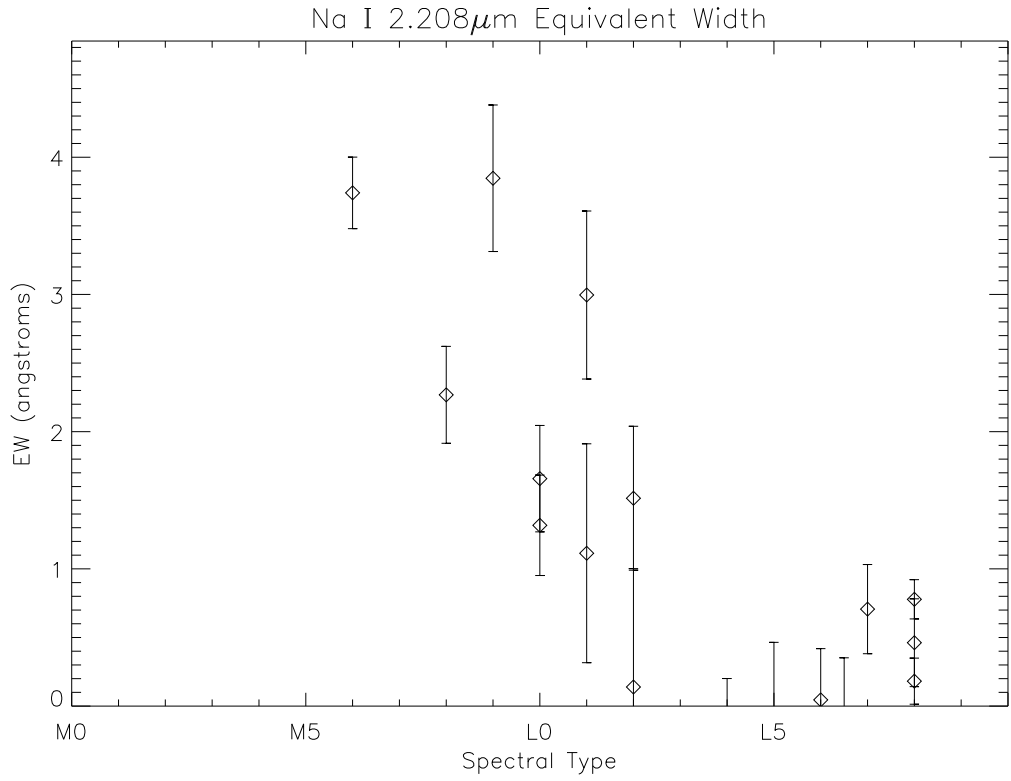} \caption{Equivalent width of the
2.2 \um doublet of Na I as a function of published spectral type
in the K99 and B02 systems. Both lines are combined. The line is
not detectable later than L2.}
\end{figure}
\epsscale{1}

\begin{figure}[!htp]
\epsscale{0.9} \plotone{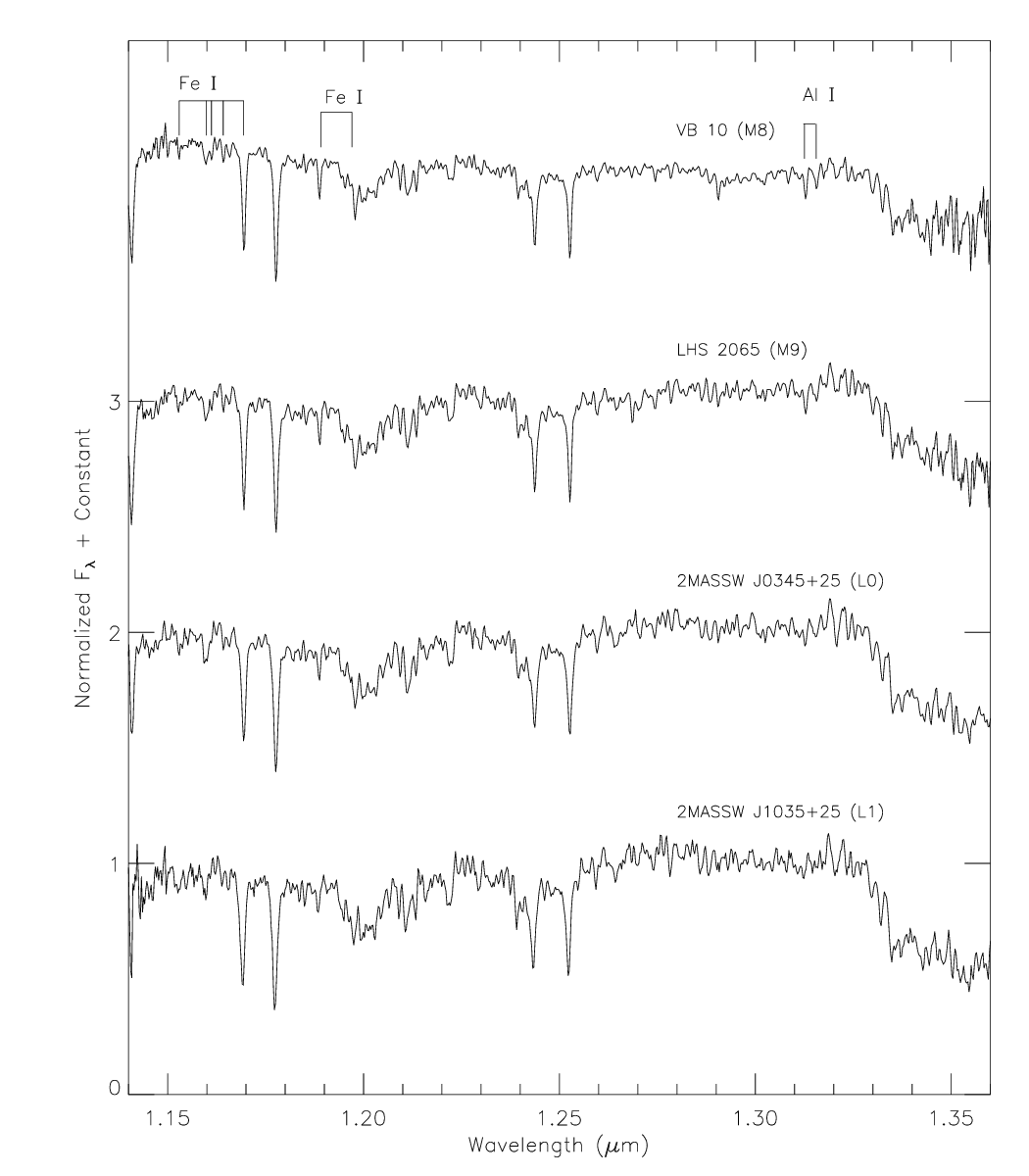} \caption{A sequence of $J$-band
spectra illustrating the change in strength of the Al I doublet at
1.314 \um and the strong Fe I line at 1.189 \um across the M to L
dwarf boundary. There is contamination by \h2o near the Al I
feature.}
\end{figure}
\epsscale{1}

\begin{figure}[!htp]
\epsscale{0.9} \plotone{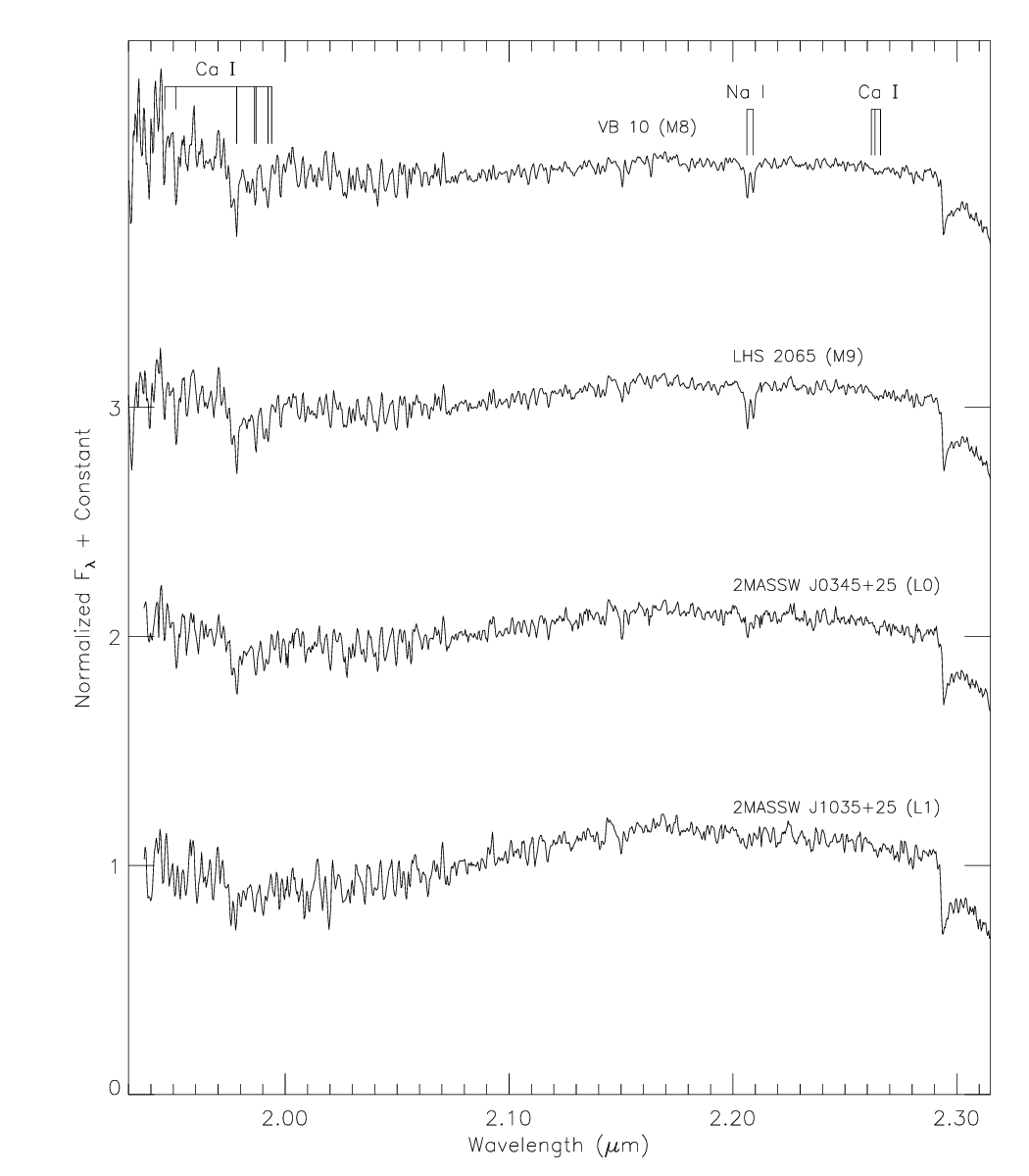} \caption{A sequence of four
$K$-band spectra with good signal-to-noise illustrating the change
in strength of the Ca I triplet at 1.98 \um and the Na I doublet
at 2.2 \um across the M to L dwarf boundary. The region containing
the Ca I line is heavily contaminated by \h2o absorption in the
source. There is relatively little change in the CO band strength
at 2.295 \um.}
\end{figure}
\epsscale{1}

\begin{figure}[!htp]
\epsscale{0.9} \plotone{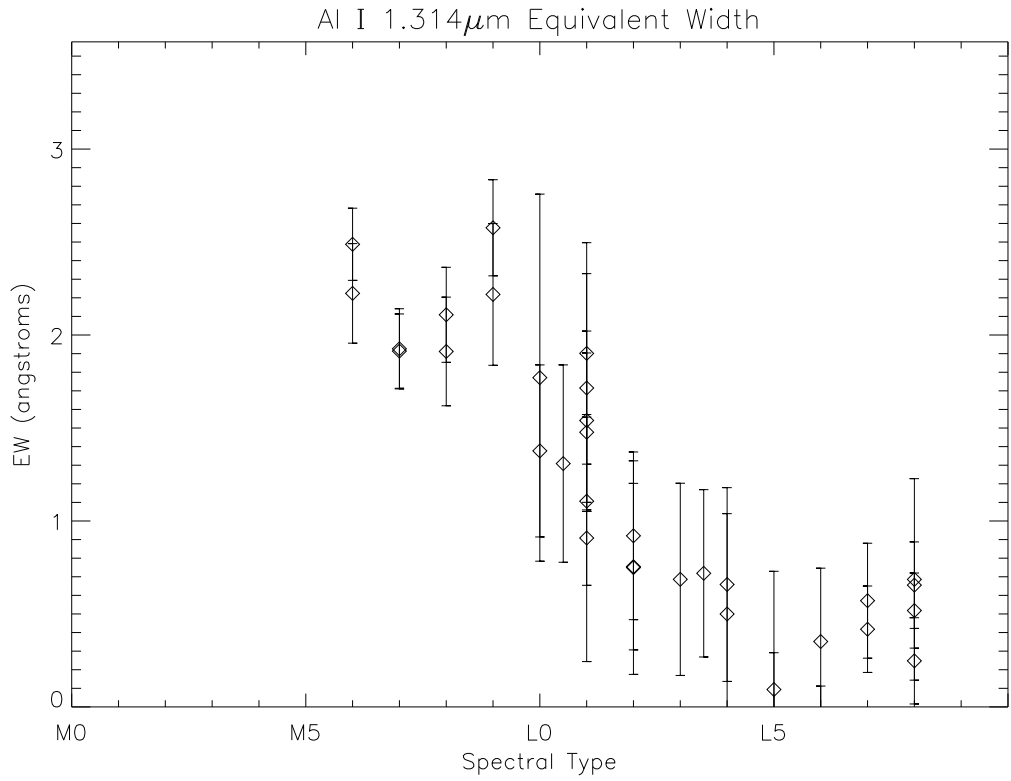} \caption{Equivalent width of the
1.314 \um doublet of Al I (in \AA) as a function of published
spectral type on the K99 and B02 system. Both lines are combined.
The line is indistinguishable from \h2o features later than L1.}
\end{figure}
\epsscale{1}

\begin{figure}[!htp]
\epsscale{0.9} \plotone{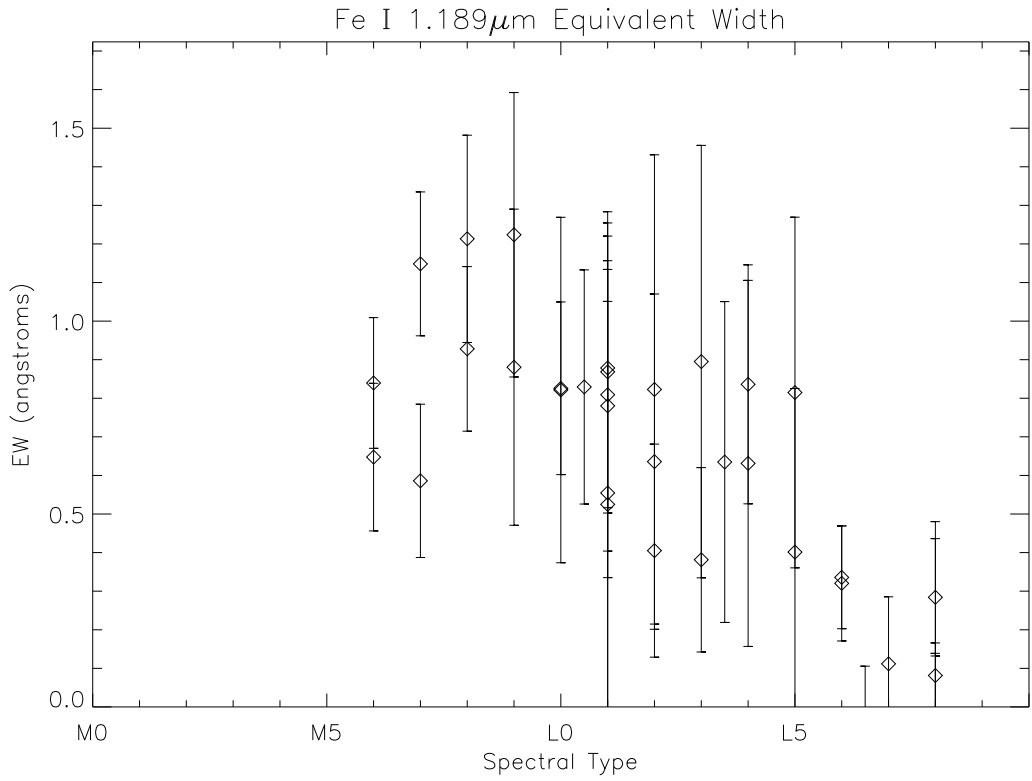} \caption{Equivalent width of the
1.189 \um line of Fe I as a function of published spectral type in
the K99 and B02 systems.}
\end{figure}
\epsscale{1}

\begin{figure}[!htp]
\epsscale{0.9} \plotone{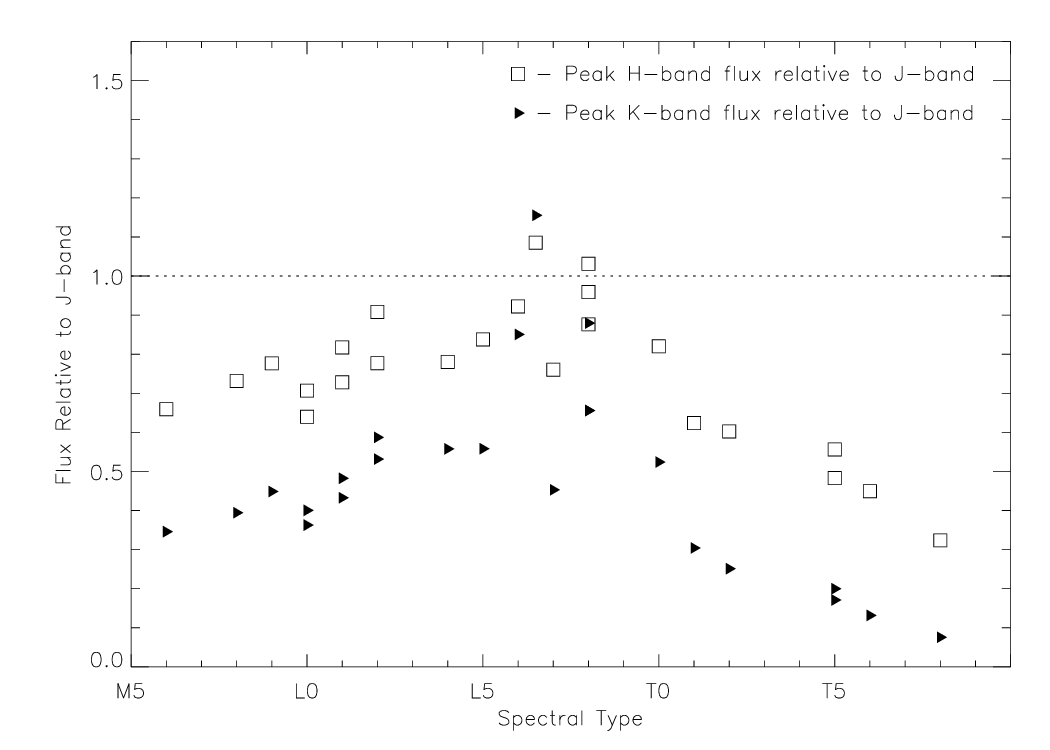} \caption{Plot of the ratio of the
peak calibrated $H$-band flux to that in the $J$-band (squares) as
a function of published spectral type. Triangles represent the
ratio of peak $K$-band flux to $J$-band. Objects above the dashed
horizontal line have $J$-band peak flux less than $H$ or $K$. T
dwarfs have $H$ and $K$ peak fluxes more depressed relative to $J$
than those of L dwarfs.}
\end{figure}
\epsscale{1}

\begin{figure}[!htp]
\epsscale{0.9} \plotone{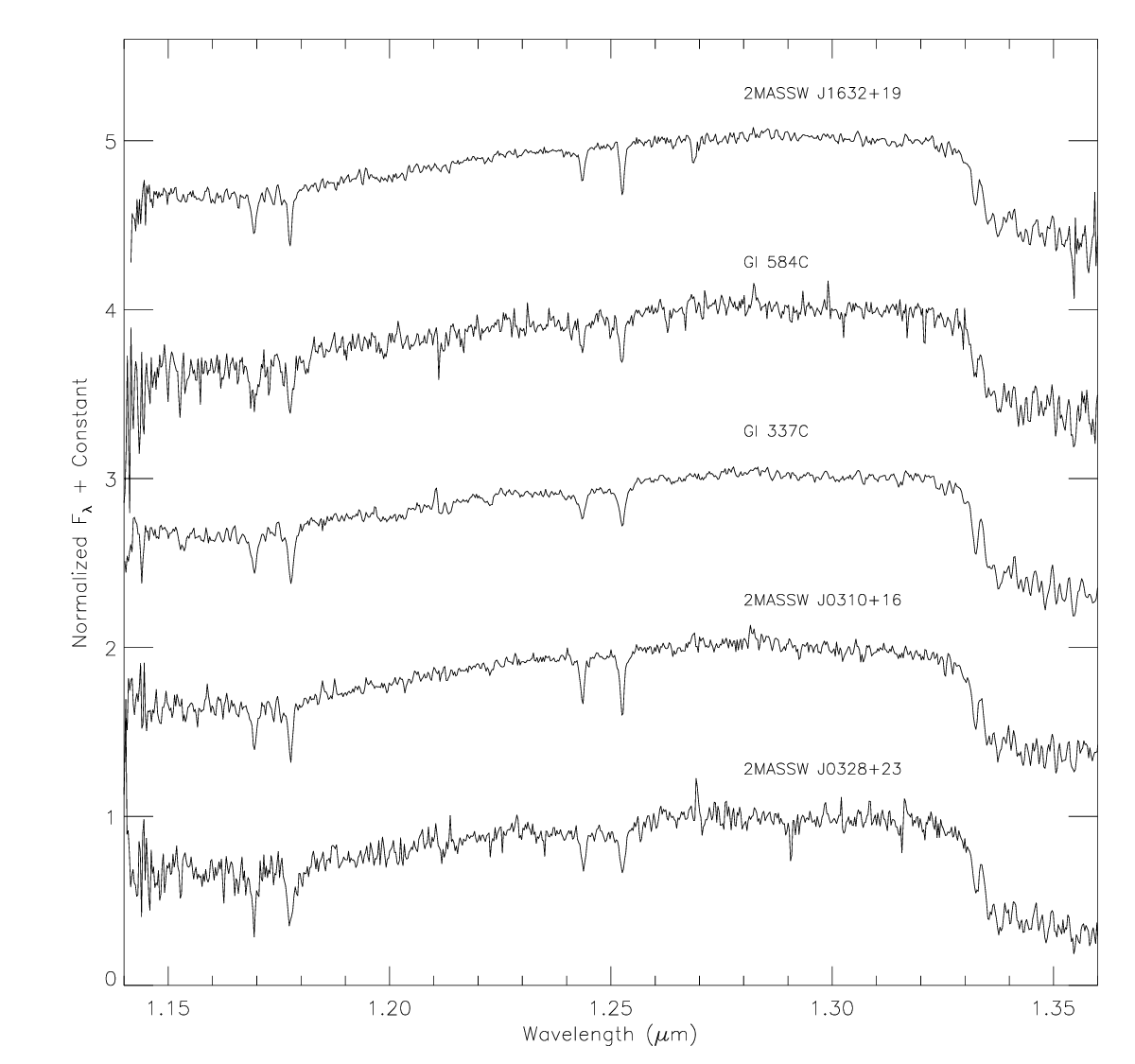} \caption{$J$-band spectra of five
L8 dwarfs are plotted to investigate diversity in the near
infrared among objects of equivalent optically-based
classification. Spectra are normalized to 1.0 at 1.265 \um and
offset by integers as in previous figures. Extended tick marks on
the vertical axis denote zero flux levels for each spectrum.}
\end{figure}
\epsscale{1}

\begin{figure}[!htp]
\epsscale{0.9} \plotone{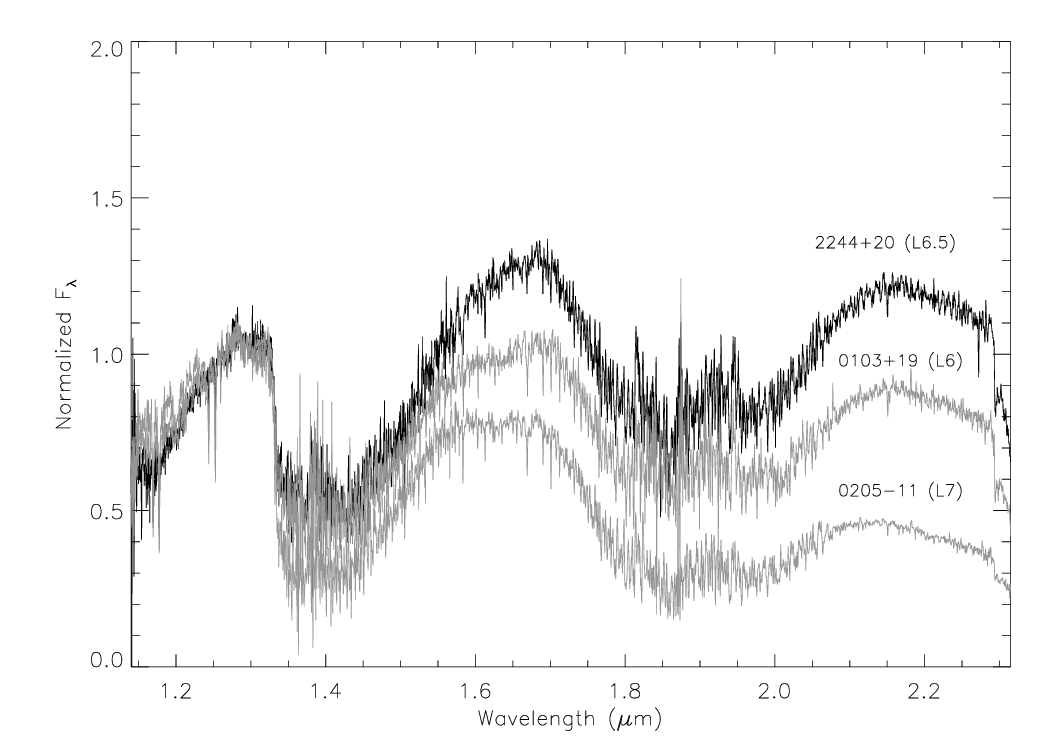} \caption{The spectrum of the
peculiar L6.5 2MASSW J2244+20 (dark solid line) is plotted and
compared to objects of similar optical spectral type, 2MASSW
J0103+19 (L6) and DENIS 0205-11 (L7). All spectra are normalized
to 1.0 at 1.270 \um. }
\end{figure}
\epsscale{1}

\begin{figure}[!htp]
\epsscale{0.9} \plotone{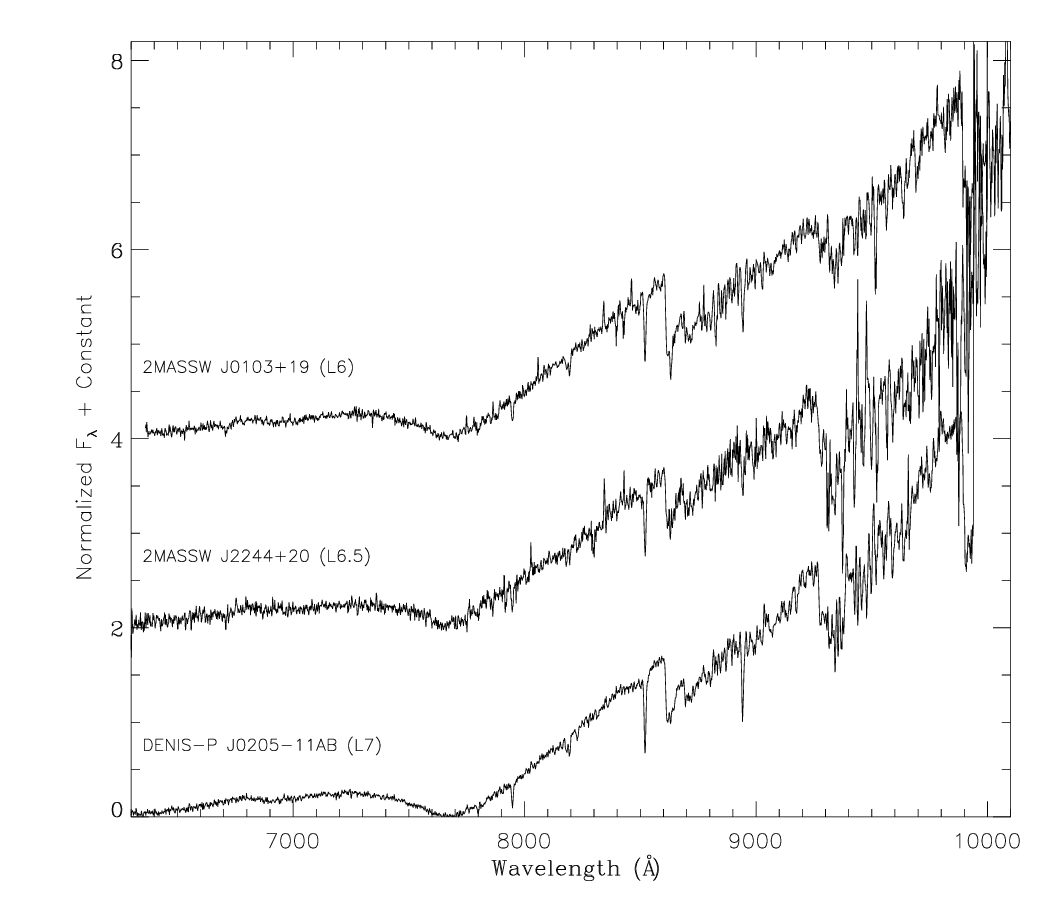} \caption{A far red optical
spectrum of the peculiar L6.5 2MASSW J2244+20 is plotted and
compared to objects of similar optical spectral type, 2MASSW
J0103+19 (L6) and DENIS 0205-11 (L7). The observations were made
by JDK using the LRIS spectrograph at the Keck Observatory.}
\end{figure}
\epsscale{1}

\clearpage

\begin{deluxetable}{lcccc}
\tabletypesize{\scriptsize}
\tablecaption{NIRSPEC Filters}
\tablewidth{0pt}
\tablehead{\colhead{Configuration} & \colhead{Filter} &
\colhead{CDG\tablenotemark{a}} &\colhead{Wavelength Range (\um)} &
\colhead{Band coverage}}
\startdata
N1 & NIRSPEC-1 & 34.95 & 0.947 - 1.121  & $Z$/$Y$ \\
N2 & NIRSPEC-2 & 36.48 & 1.089 - 1.293  & $Z$/$Y$/$J$ \\
N3 & NIRSPEC-3 & 34.08 & 1.143 - 1.375  & $J$ \\
N4 & NIRSPEC-4 & 35.58 & 1.241 - 1.593  & $H$ \\
N5 & NIRSPEC-5 & 36.72 & 1.431 - 1.808  & $H$ \\
N6a & NIRSPEC-6 & 33.48 & 1.558 - 2.000 & $H$ \\
N6b & NIRSPEC-6 & 35.18 & 1.937 - 2.315 & $K$ \\
N7 & NIRSPEC-7 & 35.66 & 1.997 - 2.428 & $K$ \\
\enddata
\tablenotetext{a}{Cross dispersed grating angle in degrees}
\end{deluxetable}

\clearpage

\begin{deluxetable}{llccccccl}
\tabletypesize{\tiny}
\tablecaption{Observing Log\tablenotemark{a}}
\tablewidth{0pt}
\tablehead{\colhead{} & \colhead{Spectral} & \colhead{R.A.} &
\colhead{Decl.} & \colhead{2MASS} & \colhead{2MASS} & \colhead{2MASS} &
\colhead{UT Date(s)} & \colhead{} \\ \colhead{Object} & \colhead{Type} &
\colhead{(J2000.0)} & \colhead{(J2000.0)} & \colhead{$J$(mag)} &
\colhead{$H$(mag)} & \colhead{$K$s(mag)} & \colhead{Observed} & \colhead{Coverage}}
\startdata
Wolf 359 (Gl 406)& M6  &  10 56 28.9  &  07 00 53   &  7.03 $\pm$0.03  &  6.48 $\pm$0.03  & 6.06 $\pm$0.03  &  2000 Dec 06  &  N3,4,6a,6b \\
                 &     &              &             &                  &                  &                 &  2001 Dec 30  &  N1,2       \\
Gl 283B          & M6  &  07 40 20.0  & -17 24 48   &  10.14$\pm$0.03  &  9.61 $\pm$0.03  & 9.26 $\pm$0.03  &  2001 Dec 31  &  N3         \\
LHS 2351         & M7  &  11 06 18.9  &  04 28 32   &  12.35$\pm$0.03  &  11.76$\pm$0.03  & 11.34$\pm$0.03  &  2000 Dec 14  &  N3 \\
VB 8             & M7  &  16 55 35.74 & -08 23 36.0 &  9.77 $\pm$0.03  &  9.19 $\pm$0.03  & 8.82 $\pm$0.03  &  2001 Jun 10  &  N3         \\
LP 412-31        & M8  &  03 20 59.6  &  18 54 23   &  11.74$\pm$0.03  &  11.04$\pm$0.03  & 10.57$\pm$0.03  &  2000 Dec 04  &  N3         \\
VB 10            & M8  &  19 16 58.10 &  05 09 11.1 &  9.90 $\pm$0.03  &  9.24 $\pm$0.03  & 8.80 $\pm$0.03  &  2001 Jun 10  &  N1,2,3,4,6a,6b\\
LHS 2065         & M9  &  08 53 36.2  & -03 29 32   &  11.19$\pm$0.03  &  10.47$\pm$0.03  & 9.97 $\pm$0.03  &  2002 Apr 22  &  N3,4,6a,6b \\
2MASSW J1239+20  & M9  &  12 39 19.4  &  20 29 52   &  14.48$\pm$0.03  &  13.68$\pm$0.03  & 13.20$\pm$0.03  &  2002 Dec 24  &  N3         \\
2MASSW J0345+25  & L0  &  03 45 43.2  &  25 40 23   &  13.99$\pm$0.03  &  13.17$\pm$0.03  & 12.67$\pm$0.03  &  2001 Dec 29  &  N3         \\
                 &         &              &             &                  &                  &                 &  2001 Dec 30  &  N1,2,4,6a,6b\\
HD 89744B        & L0      &  10 22 14.9  &  41 14 27   &  14.89$\pm$0.04  &  14.04$\pm$0.05  & 13.62$\pm$0.05  &  2001 Mar 06  &  N1,2,3     \\
                 &         &              &             &                  &                  &                 &  2001 Jun 11  &  N4         \\
                 &         &              &             &                  &                  &                 &  2002 Apr 22  &  N6a,6b     \\
2MASSW J0746+20AB& L0.5    &  07 46 42.6  &  20 00 32   &  11.74$\pm$0.03  &  11.00$\pm$0.04  & 10.49$\pm$0.03  &  2000 Apr 27  &  N3     \\
2MASSW J0208+25  & L1      &  02 08 18.3  &  25 42 53   &  14.02$\pm$0.03  &  13.11$\pm$0.04  & 12.58$\pm$0.04  &  2000 Dec 05  &  N3         \\
2MASSW J1035+25  & L1      &  10 35 24.6  &  25 07 45   &  14.70$\pm$0.04  &  13.88$\pm$0.04  & 13.28$\pm$0.04  &  2001 Dec 31  &  N3,4,6a    \\
                 &         &              &             &                  &                  &                 &  2002 Jan 01  &  N6b        \\
2MASSW J1300+19  & L1      &  13 00 42.6  &  19 12 35   &  12.71$\pm$0.02  &  12.07$\pm$0.03  & 11.61$\pm$0.03  &  2000 Apr 26  &  N3         \\
2MASSW J1439+19  & L1      &  14 39 28.4  &  19 29 15   &  12.76$\pm$0.04  &  12.05$\pm$0.04  & 11.58$\pm$0.04  &  2000 Apr 25  &  N3         \\
2MASSW J1658+70  & L1      &  16 58 03.8  &  70 27 02   &  13.29$\pm$0.02  &  12.49$\pm$0.03  & 11.93$\pm$0.02  &  2000 Jul 29  &  N3,4,6a,6b \\
2MASSW J2130-08  & L1      &  21 30 44.6  & -08 45 20   &  14.14$\pm$0.04  &  13.33$\pm$0.05  & 12.82$\pm$0.04  &  2001 Jun 11  &  N3         \\
2MASSW J0015+35  & L2      &  00 15 44.8  &  35 16 03   &  13.82$\pm$0.04  &  12.81$\pm$0.03  & 12.24$\pm$0.03  &  2000 Dec 05  &  N3         \\
                 &         &              &             &                  &                  &                 &  2001 Oct 09  &  N1         \\
                 &         &              &             &                  &                  &                 &  2001 Oct 10  &  N4,6a,6b   \\
                 &         &              &             &                  &                  &                 &  2002 Sep 01  &  N2         \\
Kelu-1           & L2      &  13 05 40.2  & -25 41 06   &  13.42$\pm$0.02  &  12.39$\pm$0.03  & 11.73$\pm$0.03  &  1999 Apr 29  &  N3,4,6a,6b \\
2MASSW J1726+15  & L2      &  17 26 00.1  &  15 38 19   &  15.65$\pm$0.07  &  14.46$\pm$0.06  & 13.64$\pm$0.05  &  2002 Sep 01  &  N3         \\
2MASSW J1506+13  & L3      &  15 06 54.4  &  13 21 06   &  13.41$\pm$0.03  &  12.41$\pm$0.03  & 11.75$\pm$0.03  &  2000 Apr 26  &  N3         \\
2MASSW J1615+35  &  L3      &  16 15 44.2  &  35 59 01   &  14.55$\pm$0.04  &  13.55$\pm$0.04  & 12.89$\pm$0.05  &  2000 Apr 26  &  N3         \\
2MASSW J0036+18  &  L3.5    &  00 36 16.2  &  18 21 10   &  12.44$\pm$0.04  &  11.58$\pm$0.03  & 11.03$\pm$0.03  &  2000 Jul 28  &  N3         \\
GD 165B          &  L4      &  14 24 39.1  &  09 17 10   &  15.55$\pm$0.06  &  14.55$\pm$0.09  & 14.06$\pm$0.07  &  1999 Jun 03  &  N3         \\
                 &          &              &             &                  &                  &                 &  2001 Jun 11  &  N1,2,4,6a,6b\\
2MASSI J2158-15  &  L4      &  21 58 04.6  & -15 50 10   &  15.04$\pm$0.05  &  13.87$\pm$0.06  & 13.19$\pm$0.05  &  2001 Oct 09  &  N3         \\
DENIS-P J1228-15AB &  L5      &  12 28 15.2  & -15 47 34   &  14.38$\pm$0.03  &  13.36$\pm$0.03  & 12.81$\pm$0.03  &  2001 Jun 03  &  N3         \\
2MASSW J1507-16  &  L5      &  15 07 47.7  & -16 27 39   &  12.82$\pm$0.03  &  11.90$\pm$0.03  & 11.30$\pm$0.03  &  1999 Jun 03  &  N3,4,6a,6b \\
2MASSW J0103+19  &  L6      &  01 03 32.0  &  19 35 36   &  16.26$\pm$0.09  &  14.88$\pm$0.06  & 14.15$\pm$0.07  &  2000 Dec 04  &  N3         \\
                 &          &              &             &                  &                  &                 &  2000 Dec 05  &  N4,6b      \\
                 &          &              &             &                  &                  &                 &  2001 Oct 09  &  N1         \\
                 &          &              &             &                  &                  &                 &  2002 Jan 01  &  N2,6a      \\
2MASSW J0850+10AB&  L6      &  08 50 35.9  &  10 57 16   &  16.46$\pm$0.12  &  15.23$\pm$0.10  & 14.46$\pm$0.07  &  2000 Dec 06  &  N3         \\
2MASSW J2244+20  &  L6.5    &  22 44 31.7  &  20 43 43   &  16.41$\pm$0.13  &  14.97$\pm$0.07  & 13.93$\pm$0.07  &  2001 Oct 10  &  N3,4,6a,6b \\
DENIS-P J0205-11AB &  L7      &  02 05 29.4  & -11 59 30   &  14.58$\pm$0.03  &  13.59$\pm$0.03  & 12.98$\pm$0.04  &  1999 Aug 20  &  N3,4,6a,6b \\
                 &          &              &             &                  &                  &                 &  2002 Sep 01  &  N1,2       \\
2MASSW J1728+39AB&  L7      &  17 28 11.5  &  39 48 59   &  15.96$\pm$0.08  &  14.78$\pm$0.07  & 13.90$\pm$0.05  &  2000 Apr 27  &  N3         \\
2MASSW J0310+16  &  L8      &  03 10 59.9  &  16 48 16   &  16.43$\pm$0.11  &  14.95$\pm$0.07  & 14.40$\pm$0.10  &  2001 Oct 09  &  N3         \\
2MASSI J0328+23  &  L8      &  03 28 42.7  &  23 02 05   &  16.67$\pm$0.14  &  15.62$\pm$0.13  & 14.84$\pm$0.13  &  2001 Dec 29  &  N3         \\
Gl 337C          &  L8      &  09 12 14.7  &  14 59 40   &  15.55$\pm$0.11  &  14.66$\pm$0.10  & 14.03$\pm$0.08  &  2001 Dec 29  &  N3         \\
                 &          &              &             &                  &                  &                 &  2001 Dec 30  &  N2         \\
                 &          &              &             &                  &                  &                 &  2002 Jan 01  &  N1,4,6a,6b \\
Gl 584C          &  L8      &  15 23 22.6  &  30 14 56   &  16.32$\pm$0.11  &  15.00$\pm$0.07  & 14.24$\pm$0.07  &  1999 Jun 03  &  N3         \\
                 &          &              &             &                  &                  &                 &  1999 Aug 20  &  N4,6a,6b   \\
2MASSW J1632+19  &  L8      &  16 32 29.1  &  19 04 41   &  15.86$\pm$0.07  &  14.59$\pm$0.05  & 13.98$\pm$0.05  &  2002 Aug 31  &  N3,4,6a,6b \\
                 &          &              &             &                  &                  &                 &  2002 Sep 01  &  N1,2       \\
SDSSP J0423-04   &  T0      &  04 23 48.6  & -04 14 04   &  14.45$\pm$0.03  &  13.44$\pm$0.04  & 12.94$\pm$0.04  &  2001 Oct 10  &  N4,6a,6b   \\
                 &          &              &             &                  &                  &                 &  2001 Dec 31  &  N1,3       \\
                 &          &              &             &                  &                  &                 &  2002 Dec 23  &  N2         \\
SDSSP J0151+12   &  T1      &  01 51 41.7  &  12 44 30   &  16.52$\pm$0.13  &  15.58$\pm$0.12  & 15.09$\pm$0.19  &  2001 Oct 09  &  N3         \\
SDSSP J0837-00   &  T1      &  08 37 17.2  & -00 00 18   &  16.77$\pm$0.21  &  $>$15.8         & $>$15.3         &  2000 Dec 05  &  N3,4,6a,6b \\
SDSSP J1254-01   &  T2      &  12 54 53.9  & -01 22 47   &  14.88$\pm$0.04  &  14.04$\pm$0.04  & 13.83$\pm$0.06  &  2001 Mar 07  &  N1,2,3,6b  \\
                 &          &              &             &                  &                  &                 &  2001 Jun 10  &  N4,6a      \\
SDSSP J1021-03   &  T3      &  10 21 09.7  & -03 04 20   &  16.26$\pm$0.10  &  15.33$\pm$0.11  & 15.10$\pm$0.18  &  2001 Jun 11  &  N3         \\
SDSSP J1750+17   &  T3.5    &  17 50 33.0  &  17 59 04   &  16.58$\pm$0.12  &  15.97$\pm$0.16  & $>$16.0         &  2001 Oct 09  &  N3         \\
SDSSP J0926+58   &  T4.5    &  09 26 15.4  &  58 47 21   &  15.72$\pm$0.06  &  15.33$\pm$0.09  & 15.44$\pm$0.17  &  2002 Dec 23  &  N3         \\
2MASSW J2254+31  &  T5      &  22 54 18.8  &  31 23 49   &  15.28$\pm$0.05  &  15.04$\pm$0.09  & 14.83$\pm$0.14  &  2000 Jul 25  &  N1,3,4     \\
                 &          &              &             &                  &                  &                 &  2000 Jul 29  &  N6a,6b     \\
                 &          &              &             &                  &                  &                 &  2002 Aug 31  &  N2         \\
2MASSW J0559-14  &  T5      &  05 59 19.1  & -14 04 48   &  13.83$\pm$0.03  &  13.68$\pm$0.04  & 13.61$\pm$0.05  &  2000 Mar 06  &  N1,2,3,4,6a,6b\\
2MASSW J2356-15  &  T6      &  23 56 54.7  & -15 53 11   &  15.80$\pm$0.06  &  15.64$\pm$0.10  & 15.83$\pm$0.19  &  1999 Aug 19  &  N3,4,6a,6b \\
                 &          &              &             &                  &                  &                 &  2001 Oct 09  &  N1         \\
                 &          &              &             &                  &                  &                 &  2002 Sep 01  &  N2         \\
SDSSP J1624+00   &  T6      &  16 24 14.4  &  00 29 15   &  15.49$\pm$0.06  &  15.52$\pm$0.10  & $>$15.4         &  1999 Jun 02  &  N3         \\
2MASSW J1237+65  &  T6.5    &  12 37 39.2  &  65 26 15   &  16.03$\pm$0.09  &  15.72$\pm$0.16  & $>$15.9         &  2000 Jul 14  &  N3         \\
2MASSW J0727+17  &  T7      &  07 27 18.2  &  17 10 01   &  15.55$\pm$0.07  &  15.82$\pm$0.18  & 15.56$\pm$0.21  &  2001 Dec 29  &  N3         \\
2MASSW J1553+15  &  T7      &  15 53 02.2  &  15 32 36   &  15.81$\pm$0.08  &  15.92$\pm$0.17  & 15.51$\pm$0.19  &  2001 Jun 11  &  N3         \\
Gl 570D          &  T8      &  14 57 15.0  & -21 21 48   &  15.33$\pm$0.05  &  15.28$\pm$0.09  & 15.27$\pm$0.17  &  2001 Mar 06  &  N1,2,3,4,6a\\
                 &          &              &             &                  &                  &                 &  2001 Mar 07  &  N6b        \\
2MASSW J0415-09  &  T8      &  04 15 19.5  & -09 35 06   &  15.71$\pm$0.06  &  15.57$\pm$0.12  & 15.45$\pm$0.20  &  2000 Dec 06  &  N3         \\
                 &          &              &             &                  &                  &                 &  2001 Mar 07  &  N5,6b      \\
                 &          &              &             &                  &                  &                 &  2001 Mar 07  &  N6b        \\

\enddata
\tablenotetext{a} {Object names are truncated for convenience. The
full 2MASS name can be derived from the given J2000 coordinates.}
\tablenotetext{-} {M dwarfs - Magnitudes from Leggett 1992 except
for 2MASSW J1239+20 (2MASS database). Spectral types based on
previously published types.}
\tablenotetext{-}{L dwarfs -
Magnitudes from 2MASS database provided by Kirkpatrick via L dwarf
archive website (See section 1).  Spectral types based on
Kirkpatrick classification scheme (K99).}
\tablenotetext{-}{T
dwarfs - Magnitudes from 2MASS database. Spectral types based on
Burgasser classification scheme (B02)}
\end{deluxetable}

\clearpage

\begin{deluxetable}{lcc}
\tabletypesize{\scriptsize} \tablecaption{Flux Normalization
Constants: F$_\lambda$ at 1.27 $\mu$m} \tablewidth{0pt}
\tablehead{\colhead{Spectral Type} & \colhead{Object} &
\colhead{$W m^{-2}$ $\mu m^{-1}$} } \startdata
M6  & Wolf 359 (Gl 406) &     5.395e-12       \\
M8  & VB 10 (LHS 474)   &     3.798e-13       \\
M9  & LHS 2065          &     1.179e-13       \\
L0  & 2MASSW J0345+25   &     8.858e-15       \\
L0  & HD 89744B         &     3.742e-15       \\
L1  & 2MASSW J1035+25   &     4.590e-15       \\
L1  & 2MASSW J1658+70   &     1.683e-14       \\
L2  & 2MASSW J0015+35   &     9.990e-15       \\
L2  & Kelu-1            &     1.458e-14       \\
L4  & GD 165B           &     2.082e-15       \\
L5  & 2MASSW J1507-16   &     2.397e-14       \\
L6  & 2MASSW J0103+19   &     1.030e-15       \\
L6.5& 2MASSW J2244+20   &     9.239e-16       \\
L7  & DENIS-P J0205-11AB&     4.972e-15       \\
L8  & Gl 337C           &     2.018e-15       \\
L8  & Gl 584C           &     1.010e-15       \\
L8  & 2MASSW J1632+19   &     1.593e-15       \\
T0  & SDSSP J0423-04    &     5.372e-15       \\
T1  & SDSSP J0837-00    &     6.147e-16       \\
T2  & SDSSP J1254-01    &     3.732e-15       \\
T5  & 2MASSW J2254+31   &     2.422e-15       \\
T5  & 2MASSW J0559-14   &     8.714e-15       \\
T6  & 2MASSW J2356-15   &     1.243e-15       \\
T8  & Gl 570D           &     1.325e-15       \\
T8  & 2MASSW J0415-09   &     7.827e-16       \\
\enddata
\end{deluxetable}

\clearpage

\begin{deluxetable}{lcc}
\tabletypesize{\scriptsize}
\tablecaption{Flux Ratio Definitions}
\tablewidth{0pt}
\tablehead{\colhead{Ratio Name} & \colhead{Flux Ratio
\tablenotemark{a}} & \colhead{Defined for}}
\startdata
H2OA    &  $<1.343>/<1.313>$  & M, L, T\\
H2OB    &  $<1.456>/<1.570>$  & M, L, T\\
H2OC    &  $<1.788>/<1.722>$  & M, L, T\\
H2OD    &  $<1.964>/<2.075>$  & M, L, T\\
CH4A    &  $<1.730>/<1.590>$  & T\\
CH4B    &  $<2.200>/<2.100>$  & T\\
CO      &  $<2.300>/<2.285>$  & M, L\\
$J$-FeH &  $<1.200>/<1.185>$  & M, L\\
$z$-FeH &  $<0.992>/<0.986>$  & M, L, T\\
\enddata
\tablenotetext{a}{Wavelengths in \um;
$<>$ is defined as the median of values located in a 0.004 \um window
centered about that wavelength.}
\end{deluxetable}

\clearpage

\begin{deluxetable}{llccccccccc}
\tabletypesize{\scriptsize}
\tablecaption{Flux Ratio Values}
\tablewidth{0pt}
\tablehead{\colhead{Spectral Type} & \colhead{Object} &
\colhead{H$_2$OA} & \colhead{H$_2$OB} & \colhead{H$_2$OC} & \colhead{H$_2$OD} &
\colhead{CH$_4$A} & \colhead{CH$_4$B} & \colhead{$z$-FeH} & \colhead{$J$-FeH} & \colhead{CO}}
\startdata
M6  & Wolf 359 (Gl 406) & 0.876  & 1.071 & 0.801 & 1.106 & 0.886 & 0.997 & 0.704 & 0.918 & 0.737  \\
M6  & Gl 283B           & 0.766  &    -- &    -- &    -- &    -- &    -- & --    & 0.941 &    --  \\
M7  & LHS 2351          & 0.777  &    -- &    -- &    -- &    -- &    -- & --    & 0.894 &    --  \\
M7  & VB 8 (LHS 429)    & 0.753  &    -- &    -- &    -- &    -- &    -- & --    & 0.901 &    --  \\
M8  & LP 412-31         & 0.791  &    -- &    -- &    -- &    -- &    -- & --    & 0.845 &    --  \\
M8  & VB 10 (LHS 474)   & 0.760  & 0.933 & 0.761 & 1.071 & 0.858 & 1.047 & 0.615 & 0.875 & 0.738  \\
M9  & LHS 2065          & 0.754  & 0.907 & 0.757 & 1.034 & 0.935 & 1.069 & --    & 0.835 & 0.693  \\
M9  & 2MASSW J1239+20   & 0.751  &    -- &    -- &    -- &    -- &    -- & --    & 0.901 &    --  \\
L0  & 2MASSW J0345+25   & 0.680  & 0.858 & 0.714 & 1.017 & 0.932 & 1.071 & 0.492 & 0.817 & 0.690  \\
L0  & HD 89744B         & 0.623  & 0.870 & 0.728 & 1.008 & 0.906 & 1.056 & 0.484 & 0.854 & 0.713  \\
L0.5& 2MASSW J0746+20AB & 0.676  &    -- &    -- &    -- &    -- &    -- & --    & 0.792 &    --  \\
L1  & 2MASSW J0208+25   & 0.639  &    -- &    -- &    -- &    -- &    -- & --    & 0.801 &    --  \\
L1  & 2MASSW J1035+25   & 0.628  & 0.781 & 0.670 & 0.934 & 0.932 & 1.090 & --    & 0.811 & 0.606  \\
L1  & 2MASSW J1300+19   & 0.568  &    -- &    -- &    -- &    -- &    -- & --    & 0.841 &    --  \\
L1  & 2MASSW J1439+19   & 0.617  &    -- &    -- &    -- &    -- &    -- & --    & 0.792 &    --  \\
L1  & 2MASSW J1658+70   & 0.681  & 0.819 & 0.699 & 0.905 & 0.906 & 1.078 & --    & 0.820 & 0.643  \\
L1  & 2MASSW J2130-08   & 0.664  &    -- &    -- &    -- &    -- &    -- & --    & 0.768 &    --  \\
L2  & 2MASSW J0015+35   & 0.671  & 0.741 & 0.707 & 0.929 & 0.953 & 1.077 & 0.455 & 0.819 & 0.606  \\
L2  & Kelu-1            & 0.608  & 0.743 & 0.665 & 0.909 & 0.938 & 1.089 & --    & 0.864 & 0.682  \\
L2  & 2MASSW J1726+15   & 0.614  &    -- &    -- &    -- &    -- &    -- & --    & 0.871 &    --  \\
L3  & 2MASSW J1506+13   & 0.571  &    -- &    -- &    -- &    -- &    -- & --    & 0.790 &    --  \\
L3  & 2MASSW J1615+35   & 0.605  &    -- &    -- &    -- &    -- &    -- & --    & 0.797 &    --  \\
L3.5& 2MASSW J0036+18   & 0.595  &    -- &    -- &    -- &    -- &    -- & --    & 0.757 &    --  \\
L4  & GD 165B           & 0.541  & 0.676 & 0.647 & 0.823 & 0.973 & 1.071 & 0.426 & 0.781 & 0.610  \\
L4  & 2MASSI J2158-15   & 0.519  &    -- &    -- &    -- &    -- &    -- & --    & 0.801 &    --  \\
L5  & DENIS-P J1228-15AB& 0.484  &    -- &    -- &    -- &    -- &    -- & --    & 0.827 &    --  \\
L5  & 2MASSW J1507-16   & 0.504  & 0.580 & 0.586 & 0.714 & 0.899 & 1.023 & --    & 0.805 & 0.673  \\
L6  & 2MASSW J0103+19   & 0.529  & 0.584 & 0.678 & 0.738 & 0.970 & 1.027 & 0.693 & 0.964 & 0.683  \\
L6  & 2MASSW J0850+10AB & 0.467  &    -- &    -- &    -- &    -- &    -- & --    & 0.964 &    --  \\
L6.5& 2MASSW J2244+20   & 0.558  & 0.564 & 0.712 & 0.730 & 0.988 & 1.032 & --    & 1.084 & 0.743  \\
L7  & DENIS-P J0205-11AB& 0.399  & 0.514 & 0.515 & 0.656 & 0.833 & 0.936 & 0.668 & 0.937 & 0.772  \\
L7  & 2MASSW J1728+39AB & 0.540  &    -- &    -- &    -- &    -- &    -- & --    & 0.973 &    --  \\
L8  & 2MASSW J0310+16   & 0.394  &    -- &    -- &    -- &    -- &    -- & --    & 1.100 &    --  \\
L8  & 2MASSI J0328+23   & 0.389  &    -- &    -- &    -- &    -- &    -- & --    & 1.047 &    --  \\
L8  & Gl 337C           & 0.400  & 0.520 & 0.526 & 0.619 & 0.804 & 0.852 & 0.788 & 1.014 & 0.734  \\
L8  & Gl 584C           & 0.434  & 0.537 & 0.552 & 0.618 & 0.840 & 0.922 & --    & 1.055 & 0.858  \\
L8  & 2MASSW J1632+19   & 0.467  & 0.533 & 0.612 & 0.632 & 0.887 & 0.927 & 0.846 & 1.054 & 0.708  \\
T0  & SDSSP J0423-04    & 0.335  & 0.435 & 0.518 & 0.641 & 0.779 & 0.868 & 0.700 & 1.020 & 0.737  \\
T1  & SDSSP J0151+12    & 0.311  &    -- &    -- &    -- &    -- &    -- & --    & 1.093 &    --  \\
T1  & SDSSP J0837-00    & 0.266  & 0.375 & 0.417 & 0.586 & 0.716 & 0.760 & --    & 1.075 & 0.800  \\
T2  & SDSSP J1254-01    & 0.196  & 0.284 & 0.369 & 0.516 & 0.624 & 0.636 & 0.844 & 1.202 & 0.773  \\
T3  & SDSSP J1021-03    & 0.155  &    -- &    -- &    -- &    -- &    -- & --    & 1.176 &    --  \\
T3.5& SDSSP J1750+17    & 0.127  &    -- &    -- &    -- &    -- &    -- & --    & 1.143 &    --  \\
T4.5& SDSSP J0926+58    & 0.081  &    -- &    -- &    -- &    -- &    -- & --    & 1.257 &    --  \\
T5  & 2MASSW J2254+31   & 0.096  & 0.230 & 0.312 & 0.412 & 0.415 & 0.315 & 0.647 & 1.219 &    --  \\
T5  & 2MASSW J0559-14   & 0.092  & 0.188 & 0.346 & 0.407 & 0.378 & 0.269 & 0.676 & 1.218 &    --  \\
T6  & 2MASSW J2356-15   & 0.076  & 0.160 & 0.323 & 0.359 & 0.280 & 0.207 & 0.724 & 1.333 &    --  \\
T6  & SDSSP J1624+00    & 0.050  &    -- &    -- &    -- &    -- &    -- & --    & 1.494 &    --  \\
T6.5& 2MASSW J1237+65   & 0.031  &    -- &    -- &    -- &    -- &    -- & --    & 1.772 &    --  \\
T7  & 2MASSW J0727+17   & 0.035  &    -- &    -- &    -- &    -- &    -- & --    & 1.798 &    --  \\
T7  & 2MASSW J1553+15   & 0.035  &    -- &    -- &    -- &    -- &    -- & --    & 1.871 &    --  \\
T8  & Gl 570D           & 0.018  & 0.079 & 0.256 & 0.215 & 0.119 & 0.061 & 1.132 & 2.231 &    --  \\
T8  & 2MASSW J0415-09   & 0.021  &    -- &    -- & 0.240 & 0.097 & 0.089 & --    & 1.931 &    --  \\
\enddata
\end{deluxetable}

\clearpage

\begin{deluxetable}{lcccccc}
\tabletypesize{\scriptsize} \tablecaption{Parameters for Best
Linear Fits: Ratio = Slope(Sp. Type) + Intercept} \tablewidth{0pt}
\tablehead{\colhead{Ratio Name} & \colhead{Data Points} &
\colhead{Slope} & \colhead{Intercept} & \colhead{R$^{2}$} &
\colhead{Stand. Dev.} & \colhead{Range}} \startdata
H20A    & 53 &    -0.0382 &       1.073  &       0.973      & 1.1  & M, L, T \\
H20B    & 24 &    -0.0436 &       1.288  &       0.991      & 0.6  & M, L, T \\
H20C    & 24 &    -0.0254 &       0.989  &       0.914      & 1.8  & M, L, T \\
H20D    & 25 &    -0.0399 &       1.377  &       0.983      & 0.8  & M, L, T \\
CH4A    & 25 &    -0.0846 &       2.489  &       0.994      & 0.2  & T \\
CH4B    & 25 &    -0.1006 &       2.853  &       0.984      & 0.4  & T \\
\enddata
\end{deluxetable}

\clearpage

\begin{deluxetable}{llcccc}
\tabletypesize{\scriptsize} \tablecaption{K I Equivalent Widths
(\AA)} \tablewidth{0pt} \tablehead{\colhead{Sp. Type} &
\colhead{Object} & \colhead{K I 1.168\um} & \colhead{K I 1.177\um}
& \colhead{K I 1.243\um} & \colhead{K I 1.254\um}} \startdata
M6  &    Wolf 359 (Gl 406) &   3.2    $\pm$   0.2    &   4.9    $\pm$   0.2    &   3.5    $\pm$   0.2    &   3.2    $\pm$   0.2    \\
M6  &    Gl 283B           &   3.1    $\pm$   0.2    &   4.9    $\pm$   0.2    &   3.3    $\pm$   0.1    &   3.0    $\pm$   0.1    \\
M7  &    LHS 2351          &   3.4    $\pm$   0.2    &   5.3    $\pm$   0.2    &   4.3    $\pm$   0.2    &   3.6    $\pm$   0.2    \\
M7  &    VB 8 (LHS 429)    &   4.4    $\pm$   0.2    &   6.4    $\pm$   0.2    &   4.6    $\pm$   0.2    &   4.6    $\pm$   0.1    \\
M8  &    LP 412-31         &   4.9    $\pm$   0.4    &   7.2    $\pm$   0.2    &   5.5    $\pm$   0.3    &   5.2    $\pm$   0.2    \\
M8  &    VB 10 (LHS 474)   &   4.5    $\pm$   0.3    &   7.1    $\pm$   0.2    &   5.1    $\pm$   0.2    &   4.6    $\pm$   0.2    \\
M9  &    LHS 2065          &   5.6    $\pm$   0.5    &   7.7    $\pm$   0.3    &   6.1    $\pm$   0.3    &   5.2    $\pm$   0.3    \\
M9  &    2MASSW J1239+20   &   4.6    $\pm$   0.4    &   7.9    $\pm$   0.6    &   5.9    $\pm$   0.3    &   5.2    $\pm$   0.4    \\
L0  &    2MASSW J0345+25   &   6.2    $\pm$   0.5    &   8.6    $\pm$   0.3    &   6.9    $\pm$   0.4    &   6.4    $\pm$   0.3    \\
L0  &    HD 89744B         &   6.6    $\pm$   0.4    &   9.0    $\pm$   0.5    &   6.8    $\pm$   0.3    &   6.1    $\pm$   0.4    \\
L0.5&    2MASSW J0746+20AB &   7.0    $\pm$   0.6    &   9.5    $\pm$   0.4    &   7.6    $\pm$   0.5    &   6.9    $\pm$   0.4    \\
L1  &    2MASSW J0208+25   &   6.6    $\pm$   0.7    &   9.4    $\pm$   0.4    &   7.3    $\pm$   0.4    &   6.5    $\pm$   0.4    \\
L1  &    2MASSW J1035+25   &   7.6    $\pm$   0.4    &   9.3    $\pm$   0.2    &   6.8    $\pm$   0.4    &   6.6    $\pm$   0.4    \\
L1  &    2MASSW J1300+19   &   9.1    $\pm$   0.4    &   12.3   $\pm$   0.5    &   8.2    $\pm$   0.5    &   8.8    $\pm$   0.4    \\
L1  &    2MASSW J1439+19   &   7.6    $\pm$   0.5    &   10.1   $\pm$   0.3    &   7.6    $\pm$   0.2    &   7.7    $\pm$   0.4    \\
L1  &    2MASSW J1658+70   &   5.8    $\pm$   0.4    &   9.6    $\pm$   0.3    &   7.3    $\pm$   0.3    &   6.6    $\pm$   0.4    \\
L1  &    2MASSW J2130-08   &   7.6    $\pm$   0.5    &   10.2   $\pm$   0.3    &   7.7    $\pm$   0.3    &   7.4    $\pm$   0.5    \\
L2  &    2MASSW J0015+35   &   7.0    $\pm$   0.4    &   9.7    $\pm$   0.2    &   7.1    $\pm$   0.3    &   6.6    $\pm$   0.5    \\
L2  &    Kelu-1            &   3.7    $\pm$   0.9    &   9.0    $\pm$   0.7    &   6.9    $\pm$   0.4    &   6.8    $\pm$   0.4    \\
L2  &    2MASSW J1726+15   &   5.3    $\pm$   0.5    &   6.5    $\pm$   0.3    &   6.7    $\pm$   0.4    &   4.8    $\pm$   0.3    \\
L3  &    2MASSW J1506+13   &   8.2    $\pm$   0.8    &   10.6   $\pm$   0.5    &   8.4    $\pm$   0.40   &   8.9    $\pm$   0.7    \\
L3  &    2MASSW J1615+35   &   8.4    $\pm$   0.7    &   10.8   $\pm$   0.5    &   8.3    $\pm$   0.50   &   7.8    $\pm$   0.6    \\
L3.5&    2MASSW J0036+18   &   9.2    $\pm$   0.4    &   12.5   $\pm$   0.4    &   8.5    $\pm$   0.5    &   9.3    $\pm$   0.5    \\
L4  &    GD 165B           &   8.3    $\pm$   0.7    &   11.0   $\pm$   0.6    &   7.3    $\pm$   0.6    &   7.9    $\pm$   0.6    \\
L4  &    2MASSI J2158-15   &   8.1    $\pm$   1.1    &   9.6    $\pm$   0.5    &   7.5    $\pm$   0.5    &   7.5    $\pm$   0.4    \\
L5  &    DENIS-P J1228-15AB&   8.9    $\pm$   0.5    &   11.2   $\pm$   0.7    &   7.2    $\pm$   0.6    &   7.3    $\pm$   0.4    \\
L5  &    2MASSW J1507-16   &   8.8    $\pm$   0.7    &   12.1   $\pm$   0.4    &   7.0    $\pm$   0.5    &   8.2    $\pm$   0.5    \\
L6  &    2MASSW J0103+19   &   5.8    $\pm$   0.3    &   6.2    $\pm$   0.4    &   4.3    $\pm$   0.3    &   5.3    $\pm$   0.4    \\
L6  &    2MASSW J0850+10AB &   6.5    $\pm$   0.4    &   9.2    $\pm$   0.5    &   4.2    $\pm$   0.4    &   5.4    $\pm$   0.4    \\
L6.5&    2MASSW J2244+20   &   3.8    $\pm$   0.4    &   5.0    $\pm$   0.4    &   1.8    $\pm$   0.1    &   2.3    $\pm$   0.2    \\
L7  &    DENIS-P J0205-11AB&   6.5    $\pm$   0.6    &   9.4    $\pm$   0.4    &   4.3    $\pm$   0.3    &   6.0    $\pm$   0.4    \\
L7  &    2MASSW J1728+39AB &   6.7    $\pm$   0.7    &   9.5    $\pm$   0.4    &   5.6    $\pm$   0.3    &   7.1    $\pm$   0.5    \\
L8  &    2MASSW J0310+16   &   5.9    $\pm$   0.6    &   8.0    $\pm$   0.5    &   3.9    $\pm$   0.1    &   5.3    $\pm$   0.1    \\
L8  &    2MASSW J0328+23   &   7.0    $\pm$   0.7    &   9.5    $\pm$   1.0    &   3.3    $\pm$   0.5    &   5.3    $\pm$   0.5    \\
L8  &    Gl 337C           &   5.8    $\pm$   0.3    &   8.0    $\pm$   0.3    &   3.2    $\pm$   0.3    &   4.8    $\pm$   0.3    \\
L8  &    Gl 584C           &   7.0    $\pm$   0.8    &   8.5    $\pm$   0.8    &   3.2    $\pm$   0.2    &   5.6    $\pm$   0.5    \\
L8  &    2MASSW J1632+19   &   4.7    $\pm$   0.3    &   6.3    $\pm$   0.4    &   3.0    $\pm$   0.1    &   3.5    $\pm$   0.2    \\
T0  &    SDSSP J0423-04    &   6.8    $\pm$   0.7    &   9.5    $\pm$   0.5    &   4.4    $\pm$   0.2    &   6.2    $\pm$   0.2    \\
T1  &    SDSSP J0151+12    &   6.1    $\pm$   0.6    &   9.6    $\pm$   0.9    &   --\tablenotemark{b}   &   6.0    $\pm$   0.3    \\
T1  &    SDSSP J0837-00    &   7.4    $\pm$   0.7    &   11.5   $\pm$   0.8    &   4.7    $\pm$   0.2    &   7.7    $\pm$   0.2    \\
T2  &    SDSSP J1254-01    &   8.2    $\pm$   1.1    &   10.0   $\pm$   0.9    &   4.7    $\pm$   0.2    &   7.7    $\pm$   0.4    \\
T3  &    SDSSP J1021-03    &   6.1    $\pm$   0.7    &   --\tablenotemark{b}   &   4.9    $\pm$   0.3    &   7.7    $\pm$   0.3    \\
T3.5&    SDSSP J1750+17    &   10.2   $\pm$   0.7    &   12.4   $\pm$   1.1    &   4.7    $\pm$   0.3    &   8.2    $\pm$   0.5    \\
T4.5&    SDSSP J0926+58    &   9.6    $\pm$   1.4    &   11.3   $\pm$   1.1    &   5.1    $\pm$   0.4    &   8.4    $\pm$   0.8    \\
T5  &    2MASSW J2254+31   &   9.5    $\pm$   0.8    &   12.0   $\pm$   0.9    &   5.1    $\pm$   0.3    &   7.7    $\pm$   0.8    \\
T5  &    2MASSW J0559-14   &   10.0   $\pm$   1.0    &   12.5   $\pm$   1.1    &   4.9    $\pm$   0.4    &   8.7    $\pm$   0.5    \\
T6  &    2MASSW J2356-15   &   9.3    $\pm$   1.9    &   12.5   $\pm$   1.2    &   5.5    $\pm$   0.6    &   9.0    $\pm$   0.5    \\
T6  &    SDSSP J1624+00    &   5.5    $\pm$   1.4    &   5.7    $\pm$   1.5    &   1.4    $\pm$   0.8    &   5.1    $\pm$   0.8    \\
T6.5&    2MASSW J1237+65   &   7.6    $\pm$   2.6    &   10.3   $\pm$   2.0    &   0.2    $\pm$   0.6    &   2.0    $\pm$   0.8    \\
T7  &    2MASSW J0727+17   &   5.3    $\pm$   2.3    &   7.1    $\pm$   1.3    &   2.3    $\pm$   0.4    &   5.2    $\pm$   0.7    \\
T7  &    2MASSW J1553+15   &   4.9    $\pm$   3.4    &   7.0    $\pm$   2.8    &   2.2    $\pm$   0.7    &   4.8    $\pm$   0.7    \\
T8  &    Gl 570D           &   --\tablenotemark{a}   &   3.6    $\pm$   1.3    &   1.7    $\pm$   0.6    &   2.6    $\pm$   0.8    \\
T8  &    2MASSW J0415-09   &   --\tablenotemark{a}   &   6.2    $\pm$   1.3    &   1.0    $\pm$   0.7    &   1.8    $\pm$   0.9    \\
\tablenotetext{a}{No detectable K I line}
\tablenotetext{b}{Erroneous data point - Spike near line center}
\enddata
\end{deluxetable}

\clearpage

\begin{deluxetable}{llcccc}
\tabletypesize{\scriptsize} \tablecaption{Na I, Al I and Fe I
Equivalent Widths (\AA)} \tablewidth{0pt}
\tablehead{\colhead{Spectral Type} & \colhead{Object} &
\colhead{Na I 1.140\um} & \colhead{Fe I 1.189\um} & \colhead{Al I
1.314\um} & \colhead{Na I 2.208\um}} \startdata
M6  &    Wolf 359 (Gl 406)      &   11.8    $\pm$   0.4 &   0.8 $\pm$   0.2 &   2.2 $\pm$   0.3 &  3.7$\pm$0.3  \\
M6  &    Gl 283B                &               &   0.6 $\pm$   0.2 &   2.5 $\pm$   0.2 & \\
M7  &    LHS 2351               &               &   1.1 $\pm$   0.2 &   1.9 $\pm$   0.2 & \\
M7  &    VB 8 (LHS 429)         &               &   0.6 $\pm$   0.2 &   1.9 $\pm$   0.2 & \\
M8  &    LP 412-31              &               &   1.2 $\pm$   0.3 &   2.1 $\pm$   0.3 & \\
M8  &    VB 10 (LHS 474)        &   13.0    $\pm$   0.7 &   0.9 $\pm$   0.2 &   1.9 $\pm$   0.3 &  2.3$\pm$0.4  \\
M9  &    LHS 2065               &               &   0.9 $\pm$   0.4 &   2.2 $\pm$   0.4 &  3.8$\pm$0.5  \\
M9  &    2MASSW J1239+20        &               &   1.2 $\pm$   0.4 &   2.6 $\pm$   0.3 & \\
L0  &    2MASSW J0345+25        &   12.3    $\pm$   0.8 &   0.8 $\pm$   0.2 &   1.4 $\pm$   0.5 &  1.3$\pm$0.4  \\
L0  &    HD 89744B              &   13.5    $\pm$   0.9 &   0.8 $\pm$   0.4 &   1.8 $\pm$   1.0 &  1.7$\pm$0.4  \\
L0.5    &    2MASSW J0746+20AB  &               &   0.8 $\pm$   0.3 &   1.3 $\pm$   0.5 & \\
L1  &    2MASSW J0208+25        &               &   0.9 $\pm$   0.4 &   1.5 $\pm$   0.5 & \\
L1  &    2MASSW J1035+25        &               &   0.5 $\pm$   0.5 &   1.1 $\pm$   0.5 &  1.1$\pm$0.8  \\
L1  &    2MASSW J1300+19        &               &   0.8 $\pm$   0.4 &   1.7 $\pm$   0.6 & \\
L1  &    2MASSW J1439+19        &                   &   0.8 $\pm$   0.5 &   0.9 $\pm$   0.7 & \\
L1  &    2MASSW J1658+70        &               &   0.9 $\pm$   0.4 &   1.9 $\pm$   0.6 &  3.0$\pm$0.6  \\
L1  &    2MASSW J2130-08        &               &   0.6 $\pm$   0.6 &   1.5 $\pm$   0.4 & \\
L2  &    2MASSW J0015+35        &   13.3    $\pm$   1.1 &   0.6 $\pm$   0.4 &   0.9 $\pm$   0.5 &  1.5$\pm$0.5  \\
L2  &    Kelu-1                 &               &   0.8 $\pm$   0.6 &   0.7 $\pm$   0.6 &  0.1$\pm$0.9  \\
L2  &    2MASSW J1726+15        &               &   0.4 $\pm$   0.3 &   0.8 $\pm$   0.4 & \\
L3  &    2MASSW J1506+13        &               &   0.4 $\pm$   0.2 &   0.7 $\pm$   0.5 & \\
L3  &    2MASSW J1615+35        &               &   0.9 $\pm$   0.6 &   1.9 $\pm$   0.4 & \\
L3.5    &    2MASSW J0036+18        &                   &   0.6 $\pm$   0.4 &   0.7 $\pm$   0.4 & \\
L4  &    GD 165B                &   14.5    $\pm$   0.7 &   0.6 $\pm$   0.5 &   0.7 $\pm$   0.5 &  -0.1$\pm$0.3  \\
L4  &    2MASSI J2158-15        &               &   0.8 $\pm$   0.3 &   0.5 $\pm$   0.5 & \\
L5  &    DENIS-P J1228-15AB &               &   0.8 $\pm$   0.5 &   0.1 $\pm$   0.6 & \\
L5  &    2MASSW J1507-16        &               &   0.4 $\pm$   0.4 &   -0.1    $\pm$   0.4 &  -0.1$\pm$0.6  \\
L6  &    2MASSW J0103+19        &   7.1 $\pm$   1.7 &   0.3 $\pm$   0.1 &   -0.2    $\pm$   0.3 &  0.0$\pm$0.4  \\
L6  &    2MASSW J0850+10AB  &               &   0.3 $\pm$   0.1 &   0.4 $\pm$   0.4 & \\
L6.5    &    2MASSW J2244+20    &                   &   -0.2    $\pm$   0.3 &  --\tablenotemark{a} &  -0.1$\pm$0.5  \\
L7  &    DENIS-P J0205-11AB &   7.9 $\pm$   1.1 &   0.1 $\pm$   0.2 &   0.6 $\pm$   0.3 &  0.7$\pm$0.3  \\
L7  &    2MASSW J1728+39AB      &               &   -0.1    $\pm$   0.1 &   0.4 $\pm$   0.2 & \\
L8  &    2MASSW J0310+16        &               &   0.1 $\pm$   0.4 &   --\tablenotemark{a}   & \\
L8  &    2MASSW J0328+23        &               &   -0.5    $\pm$   0.6 &   0.7 $\pm$   0.5 & \\
L8  &    Gl 337C                &               &   -0.2    $\pm$   0.1 &   0.5 $\pm$   0.2 &  0.8$\pm$0.1  \\
L8  &    Gl 584C                &               &   -0.3    $\pm$   0.4 &   0.2 $\pm$   0.2 &  0.5$\pm$0.3  \\
L8  &    2MASSW J1632+19        &   5.5 $\pm$   0.8 &   0.3 $\pm$   0.2 &   0.7 $\pm$   0.2 &  0.2$\pm$0.2  \\

\tablenotetext{a}{No detectable line}
\enddata
\end{deluxetable}

\clearpage

\begin{references}

Ackerman, A. S., \& Marley, M. S. 2001, ApJ, 556, 872 \\
Allard, F., Hauschildt, P. H., Alexander, D. R., Tamanai, A., \& Schweitzer, A. 2001, ApJ, 556, 357 \\
Becklin, E.E., \& Zuckerman, B. 1988, Nature, 336, 656 \\
Bergeron, P.,Wesemael, F., \& Beauchamp, A. 1995, PASP, 107, 1047\\
Burgasser, A. J., Kirkpatrick, J. D., Reid, I. N., Liebert, J., Gizis, J. E., \& Brown, M. E. 2000a, AJ, 120, 473 \\
Burgasser, A. J., Marley, M. S., Ackerman, A. S., Saumon, D., Lodders, K., Dahn, C. C., Harris, H. C., \& Kirkpatrick, J. D., 2002a, ApJ, 571, L151 \\
Burgasser, A. J., et al. 1999, ApJ, 522, L65 \\
Burgasser, A. J., et al. 2000b, ApJ, 531, L57 \\
Burgasser, A. J., et al. 2002b, ApJ, 564, 421 (B02) \\
Burrows, A., Hubbard, W. B., Lunine, J. I., \& Liebert, J. 2001, Rev. Mod. Phys., 73, 719 \\
Burrows, A., Marley, M.S., \& Sharp, C.M. 2000, ApJ, 531, 438 \\
Burrows, A., Ram, R. S., Bernath, P., Sharp, C. M., \& Milsom, J. A. 2002, ApJ, 577, 986 \\
Cannon, A. J., \& Pickering, E.C. 1901, Ann. Astron. Obs. Harvard Coll., 28, 129 \\
Cohen, M., Walker, R.G., Barlow, M.J., \& Deacon, J.R. 1992, AJ, 104, 1650\\
Cox. A.N., ed. 2000, Allen's Astrophysical Quantities (4th ed,; New York: Springer)\\
Cushing, M. C., Rayner, J. T., Davis, S. P., \& Vacca, W. D. 2003, ApJ, 582, 1066 \\
Dahn, C. C., et al. 2002, AJ, 124, 1170 \\
Delfosse, X., et al. 1997, A \& A, 327, L25\\
Dulick, M., Bauschlicher, C.W., Burrows, A., Sharp, C.M., Ram, R.S., \& Bernath, P. 2003, astro-ph/0305162\\
Epchtein, N., et al. 1997, Messenger, 87, 27 \\
Geballe, T. R., et al. 2002, ApJ, 564, 466 (G02)\\
Hillenbrand, L. A., Foster, J. B., Persson, S. E., \& Matthews, K. 2002, PASP, 114, 708 \\
Jones, H. R. A., Longmore, A. J., Jameson, R. F., \& Mountain, C. M. 1994, MNRAS, 267, 413 \\
Kirkpatrick, J. D., Beichman, C. A., \& Skrutskie, M. F. 1997, ApJ, 476, 311 \\
Kirkpatrick, J. D., Dahn, C. C., Monet, D. G., Reid, I. N., Gizis, J. E., Liebert, J., \& Burgasser, A. J. 2001, AJ, 121, 3235 \\
Kirkpatrick, J.D., Henry, T.J., \& Liebert, J. 1993, ApJ, 406, 701 \\
Kirkpatrick, J.D., et al. 1999, ApJ, 519, 802 (K99)\\
Kirkpatrick, J. D., Reid, I. N., Liebert, J., Gizis, J. E., Burgasser, A. J., Monet, D. G., Dahn, C. C., Nelson, B., \& Williams, R. J. 2000, AJ, 120, 447\\
Leggett, S. K. 1992, ApJS, 82, 351 \\
Leggett, S. K., Toomey, D. W., Geballe, T. R., \& Brown, R. H. 1999, ApJ, 517, L139 \\
Leggett, S. K., et al. 2000, ApJ, 536, L35 \\
Leggett, S. K., Allard, F., Geballe, T., Hauschildt, P. H., \& Schweitzer, A. 2001, ApJ, 548, 908 \\
Liebert, J., Reid, I.N., Burrows, A., Burgasser, A.J., Kirkpatrick, J.D., \& Gizis, J.E. 2000, ApJ, 533, L155 \\
Lodders, K. 2002, ApJ, 577, 974\\
Mart{\'{\i}}n, E., Delfosse, X., Basri, G., Goldman, B., Forveille, T., \& Zapatero Osorio, M.R. 1999, AJ, 118, 2466 \\
McGovern, M., McLean, I. S., Kirkpatrick, J. D., Burgasser, A. J., \& Prato, L. 2003, in prep.\\
McLean, I.S., et al. 1998, Proc. SPIE, 3354, 566 \\
McLean, I.S., et al. 2000a, Proc. SPIE, 4008, 1048 \\
McLean, I.S., et al. 2000b, ApJ, 533, L45 \\
McLean, I.S., et al. 2001, ApJ, 561, L115 \\
McLean, I.S., et al. 2003a, in IAU Symp. 211, Brown Dwarfs, ed. E. Martin, (San Francisco: ASP), 385\\
McLean, I.S., et al. 2003b, in prep.\\
Morgan, W.W., Keenan, P.C., \& Kellman, E. 1943, An Atlas of Stellar Spectra, with an Outline of Spectral Classification (Chicago: Univ. Chicago Press) \\
Nakajima, T., Oppenheimer, B. R., Kulkarni, S. R., Golimowski, D. A., Matthews, K., \& Durrance, S. T. 1995, Nature, 378, 463 \\
Nakajima, T., Tsuji, T., \& Yanagisawa, K. 2001, ApJ, 561, L119 \\
Oppenheimer, B. R., Kulkarni, S. R., Matthews, K., \& Nakajima, T. 1995, Science, 270, 1478 \\
Oppenheimer, B. R., Kulkarni, S. R., Matthews, K., \& van Kerkwijk, M. H. 1998, ApJ, 502, 932 \\
Pickering, E.C. 1890, Havard Coll. Obs. Ann., 27, 1 \\
Prato, L., et al. 2003, in prep.\\
Reid, I. N., Kirkpatrick, J. D., Gizis, J. E., Dahn, C. C., Monet, D. G., Williams, R. J., Liebert, J., \& Burgasser, A. J. 2000, AJ, 119, 369 \\
Reid, I. N., Burgasser, A. J., Cruz, K., Kirkpatrick, J. D., \& Gizis, J. E. 2001, AJ, 121, 1710 \\
Ruiz, M. T., Leggett, S. K., \& Allard, F. 1997, ApJ, 491, L107 \\
Saumon, D., Bergeron, P., Lunine, J. I., Hubbard, W. B., \& Burrows, A. 1994, ApJ, 424, 333 \\
Saumon, D., Geballe, T.R., Leggett, S. K., Marley, M. S.,
Freedman, R. S., Lodders, K., Fegley, B., \& Sengupta, S. K. 2000, ApJ, 541, 374 \\
Skrutskie, M. F., et al. 1997, The Impact of Large-Scale Near-IR Sky Surveys, ed. F. Garzon et al. (Dordrecht: Kluwer), 25 \\
Strauss, M.A., et al. 1999, ApJ, 522, L61\\
Testi, L., et al. 2001, ApJ, 522, L147 \\
Tokunaga, A. T. 2000, in Allen's Astrophysical Quantities, ed. A. N. Cox (4th ed.; New York: Springer), 151 \\
Tokunaga, A. T., \& Kobayashi, N. 1999, AJ, 117, 1010 \\
Tsuji, T. 2001, in Ultracool Dwarfs: New Spectral Types L and T, ed. H. R. A. Jones \& I. A. Steele (Berlin: Springer), 9 \\
Wilson, J. C., Kirkpatrick, J. D., Gizis, J. E., Skrutskie, M. F., Monet, D. G., \& Houck, J. R. 2001, AJ, 122, 1989 \\
Wing, R.F. \& Ford, W.K., Jr. 1969, PASP, 81, 527\\
York, D.G., et al. 2000, AJ, 120, 1579 \\

\end{references}
\end{document}